\documentclass[iop,twocolappendix,numberedappendix]{emulateapj}
\usepackage{epsfig}
\usepackage{amsmath}
\usepackage{enumerate}
    
\newcommand{\beq}{\begin{equation}}
\newcommand{\eeq}{\end{equation}}
\newcommand{\bea}{\begin{eqnarray}}
\newcommand{\eea}{\end{eqnarray}}
\newcommand{\non}{\nonumber \\}
\newcommand{\trm}[1]{\textrm{#1}}

\renewcommand{\vec}[1]{\mbox{\boldmath $\displaystyle #1$}}

\newcommand{\grad}{{\mbox{\boldmath $\nabla$}}}

\def\co{\!:\!}

\begin{document}

\shortauthors{WEINBERG, ARRAS, \& BURKART}
\shorttitle{NONLINEAR COUPLING OF \lowercase{$p$}-MODES TO \lowercase{$g$}-MODES}

\title{An instability due to the nonlinear coupling of \lowercase{$p$}-modes to \lowercase{$g$}-modes:\\  Implications for coalescing neutron star binaries}

\author{Nevin N.~Weinberg$^{1}$, Phil~Arras$^{2}$,  and Joshua~Burkart$^{3}$}
\affil{$^1$Department of Physics, and Kavli Institute for Astrophysics and Space Research, Massachusetts Institute of Technology, \\Cambridge, MA 02139, USA; nevin@mit.edu}
\affil{$^2$Department of Astronomy, University of Virginia, P.O. Box 
400325, Charlottesville, VA 22904-4325, USA}
\affil{$^3$Department of Physics, 366 LeConte Hall, University of California, Berkeley, CA 94720, USA}
   
\begin{abstract}
A weakly nonlinear fluid wave propagating within a star can be unstable to three-wave interactions. The resonant parametric instability is a well-known form of three-wave interaction in which a primary wave of frequency $\omega_a$ excites a pair of secondary waves of frequency $\omega_b+\omega_c\simeq \omega_a$. Here we consider a \emph{nonresonant} form of three-wave interaction in which a low-frequency primary wave excites a high-frequency $p$-mode and a low-frequency $g$-mode such that $\omega_b+\omega_c\gg \omega_a$. We show that a $p$-mode can couple so strongly to a $g$-mode of similar radial wavelength that this type of nonresonant interaction is unstable even if the primary wave amplitude is small. As an application, we analyze the stability of the tide in coalescing neutron star binaries to $p$-$g$ mode coupling.  We find that the equilibrium tide and dynamical tide are both $p$-$g$ unstable at gravitational wave frequencies $f_{\rm gw}\ga 20\trm{ Hz}$ and drive short wavelength $p$-$g$ mode pairs to significant energies on very short timescales (much less than the orbital decay time due to gravitational radiation). Resonant parametric coupling to the tide is, by contrast, either stable or drives modes at a much smaller rate.  We do not solve for the saturation of the $p$-$g$ instability and therefore we cannot say precisely how it influences the evolution of neutron star binaries.  However, we show that if even a single daughter mode saturates near its wave breaking amplitude, the $p$-$g$ instability of the equilibrium tide will: (i) induce significant orbital phase errors ($\Delta \phi \ga 1\trm{ radian}$) that accumulate primarily at low frequencies ($f_{\rm gw}\la50\trm{ Hz}$) and (ii) heat the neutron star core to a temperature of $T\sim 10^{10}\trm{ K}$. Since there are at least $\sim 100$ unstable $p$-$g$ daughter pairs, $\Delta \phi$ and $T$ are potentially much larger than these values. Tides might therefore significantly influence the gravitational wave signal and electromagnetic emission from coalescing neutron star binaries at much larger orbital separations than previously thought.
\end{abstract}


\section{Introduction}
\label{sec:intro} 

In the standard theory of stellar oscillations, one solves the linearized fluid equations with the assumption that the waves that propagate within a star do not interact with each other.  However, in some stars  internal waves are driven to such large amplitudes that the linear approximation becomes invalid. Examples of systems in which nonlinear wave interactions can be important include close binaries \citep{Kumar:96, Goodman:98, Barker:10, Weinberg:12, Fuller:12, Burkart:12}, the sun \citep{Kumar:89}, white dwarfs \citep{Wu:01}, RR Lyrae \citep{Molnar:12}, and neutron stars that are either newly formed \citep{Weinberg:08} or rapidly rotating \citep{Arras:03}.  

As long as the nonlinearities are not too strong, the wave interactions can be described using a perturbative approach. At the lowest nonlinear order, the interactions involve three-wave couplings in which a large amplitude parent wave $a$ excites pairs of daughter waves $(b,c)$.  These interactions often occur as a resonant parametric instability in which the parent's (possibly driven) oscillation frequency $\omega_a$ nearly equals the sum of the daughters'  natural oscillation frequency $\omega_b + \omega_c$ (see e.g., \citealt{Hasselmann:67}).  The studies cited above exclusively considered such resonant interactions between parents and daughters. Here we consider the stability of \emph{nonresonant} interactions among strongly coupled waves. 

The nonlinear coupling strength is sensitive to the spatial structure of the waves.  For three-wave interactions, the coupling strength is parametrized by the coupling coefficient $\kappa_{abc}$. The magnitude of $\kappa_{abc}$ is largest in regions where the radial wavenumbers satisfy $|k_b-k_c|\la k_a$, the usual condition for momentum conservation; otherwise, the waves are incoherent and their spatial oscillations tend to cancel out the interaction (\citealt{Wu:01, Weinberg:12}, hereafter WAQB). Thus, $\kappa_{abc}$ can be large for even long wavelength parents if $k_b\simeq k_c$.  For waves of high radial order and low angular degree, the dispersion relation is $k\sim \omega/c_s$ for high-frequency acoustic waves ($p$-modes) and $k\sim \Lambda N /\omega r$ for low-frequency internal gravity waves ($g$-modes; here $c_s$ is the adiabatic sound speed, $\Lambda^2=\ell(\ell+1)$, $\ell$ is the angular degree, $N$ is the Brunt-V\"ais\"al\"a frequency, and $r$ is the radial coordinate). There are therefore three ways to satisfy $k_b\simeq k_c$ at a given radius: (i) the daughters are both $g$-modes and $\Lambda_c\omega_b\simeq \Lambda_b\omega_c$, (ii) the daughters are  both $p$-modes and $\omega_b\simeq\omega_c$, or (iii) one daughter is a $p$-mode and the other is a $g$-mode and 
\beq
\label{eq:kb_eq_kc}
\omega_b \omega_c \simeq \Lambda_c N c_s /r,
\eeq
where we took $b$ to be the $p$-mode and $c$ to be the $g$-mode. In cases (i) and (ii), the three waves can satisfy the nonlinear resonance condition if $\omega_b\simeq\omega_c\simeq \omega_a/2$.  In case (iii),  $\omega_b \gg \omega_c$ and the resonance condition cannot be satisfied (assuming that $\omega_a < \omega_b$; e.g., the parent is a $g$-mode or linearly driven by a tide).   We will show that $p$-modes can couple so strongly to $g$-modes that such nonresonant interactions can, nonetheless, be unstable even for relatively small amplitude parent waves. 

The primary application of $p$-$g$ mode coupling that we consider in this paper is in the context of tides in coalescing neutron star-neutron star (NS-NS) and neutron star-black hole (NS-BH) binaries. These binaries are the most promising sources for ground-based gravitational wave observatories such as LIGO and Virgo \citep{Cutler:02}. Within the next few years, advanced versions of these detectors should be taking data with sufficient sensitivity that they will detect the first gravitational wave signature of a compact-binary coalescence \citep{LIGO:10}. Extracting the signal from detector noise requires accurate theoretical templates and thus a precise understanding of the gravitational waveform.  If a phase error of $\ga 1\trm{ radian}$ accumulates over the final $\approx 10^4$ orbits, it can lead to substantial errors in the measurement of the binary parameters or even significantly decrease the source detectability \citep{Cutler:93}. 

Tidal interactions extract energy (and angular momentum) from the orbit and, depending on the nature and rate of the internal dissipation, deposit it within the star as some combination of mode and thermal energy (and mode and spin angular momentum). As a result, tidal interactions in NS binaries modify the rate of inspiral and lead to phase shifts in the gravitational waveform that may affect source detectability if sufficiently large and unaccounted for \citep{Bildsten:92, Kochanek:92}. Conversely, tide-induced phase shifts may encode highly sought information about the NS equation of state \citep{Flanagan:08, Hinderer:10, Damour:12}. 

As the NS inspirals, tides induce a large scale distortion of the star (referred to as the equilibrium tide) and also excite resonant oscillation modes  (the dynamical tide). Although the equilibrium tide stores considerable energy,  if its dissipation is determined entirely by linear processes, it does not significantly affect the orbit (\citealt{Bildsten:92, Lai:94}).  The dynamical tide was first studied in non-rotating NSs, where the resonant modes are $g$-modes with frequency $\la 100 \trm{ Hz}$ \citep{Reisenegger:94b, Lai:94}. These studies found that while the effect on the gravitational waveform is small, linear dissipation of 
the excited $g$-modes can heat the NS core to a temperature of $\sim 10^8\trm{ K}$. 
Furthermore, rapid rotation can strongly enhance the tidal effects
 and lead to the excitation of $r$-modes and inertial waves,
resulting in phase shifts of $\sim 0.1$ to $\gg 1$ radians \citep{Ho:99, Lai:06, Flanagan:07}. However, the required spin frequencies are higher than is thought to be likely for NS-NS binaries. 

All of these studies ignored nonlinear interactions and instead assumed 
that linear theory is valid at gravitational wave frequencies $f _{\rm gw} \la 400\trm{ Hz}$.\footnote{While
several groups now carry out hydrodynamic numerical simulations of
compact object inspiral using realistic equations of state (e.g.,
\citealt{Oechslin:07, Sekiguchi:11}), they only simulate the last few
orbits before the merger ($f_{\rm gw} \ga 400\trm{ Hz}$).} Some of the studies argue that because the amplitude of the tidal perturbations are $\la 1\%$ of the NS radius at these frequencies, the linear approximation should be valid. However, the validity of the linear approximation depends on more than just the amplitude of the perturbations; it also depends on the strength of the nonlinear coupling $\kappa_{abc}$ between the primary perturbation and other modes within the star, and the individual properties of those modes such as their frequency and linear damping rates. As a result, even if the perturbation amplitude is $\ll 1\%$ of the stellar radius, it is potentially unstable to nonlinear wave interactions. Indeed, we will show that by the time a binary first enters LIGO's bandpass ($f_{\rm gw} \approx 20\trm{ Hz}$), the equilibrium and dynamical tides are unstable to nonlinear three-wave interactions. These instabilities drive rapidly growing, short wavelength modes and can potentially lead to significantly enhanced tidal dissipation relative to linear theory predictions.

The structure of the paper is as follows. In \S~\ref{sec:stability} we carry out a stability analysis of $p$-$g$ mode coupling for a linearly driven parent wave. In \S~\ref{sec:strength} we calculate the strength of $p$-$g$ mode coupling in NSs.  In \S~\ref{sec:growth_rates} we show that coalescing NS binaries  are subject to the $p$-$g$ mode coupling instability (PGI), with the equilibrium tide and dynamical tide serving as parent waves that drive the daughters.  We also evaluate the stability of the tide to the resonant parametric instability. In \S~\ref{sec:saturation} we estimate how the PGI might influence the orbit and tidal heating of coalescing NS binaries. We conclude in \S~\ref{sec:conclude} and briefly discuss the potential influence of the PGI in systems other than NS binaries.

\section{Stability of \lowercase{$p$}-\lowercase{$g$} mode coupling}
\label{sec:stability}

Consider a linearly driven parent $a$ that is coupled to a daughter pair $(b,c)$. To lowest nonlinear order, the amplitude equations take the form (see, e.g., \citealt{Schenk:02} and WAQB)
\bea
\label{eq:amp_eqn1}
\ddot{q}_a + \gamma_a\dot{q}_a+ \omega_a^2 q_a &=& \omega_a^2 \left[U_a(t) + \kappa_{abc}^\ast q_b^\ast q_c^\ast\right]\\
\label{eq:amp_eqn2}
\ddot{q}_b + \gamma_b\dot{q}_b+ \omega_b^2 q_b &=& \omega_b^2\kappa_{abc}^\ast q_a^\ast q_c^\ast\\
\label{eq:amp_eqn3}
\ddot{q}_c + \gamma_c\dot{q}_c+ \omega_c^2 q_c &=& \omega_c^2\kappa_{abc}^\ast q_a^\ast q_b^\ast,
\eea
where the $q$'s are complex mode amplitudes, the $\gamma$'s are linear damping rates, and we assume that the parent is harmonically driven as $U_a(t)=U_a e^{-i\omega t}$.  In order to determine whether the parent is stable at its linear amplitude, let $q_b=A_b e^{(s+i\sigma_b)t}$ and similarly for $q_c$, where $s$, $\sigma_b$, and $\sigma_c$ are real constants. If $\sigma_b+\sigma_c=\omega$, the harmonic time dependences cancel in the amplitude equations.  For daughters that are \emph{not} resonant with the parent ($\omega_b\gg \omega_c, \omega$),  we show in Appendix \ref{sec:app:stability_analysis} that if
\beq
\label{eq:stability_criterion}
\left|q_a\right| \ga \left|\kappa_{abc}\right|^{-1}
\eeq
 the daughters are unstable ($s>0$) and grow exponentially at a rate
\beq
\label{eq:growth_rate}
\Gamma_{bc} \simeq \omega_c \left|\kappa_{abc} q_a\right|.
\eeq 
The stability criterion applies even if the linear damping rates are large, i.e., $\gamma_{b,c}\sim \omega_{b,c}$. Furthermore, even a static ($\omega=0$)  parent can be unstable.\footnote{The instability of an $\omega=0$ parent implies that the static background is unstable to small perturbations (as in a Rayleigh-Taylor instability).  Note, however, that we do not consider $\omega=0$ parents in this paper; even the equilibrium tide is time-dependent (e.g., the $\ell=2, m=\pm2$ harmonic oscillates at $f_{\rm gw}$).} Such an instability has the features of what \citet{Wu:98} refer to as an amplitude instability.

\begin{figure}[t!]
\epsscale{1.1}
\plotone{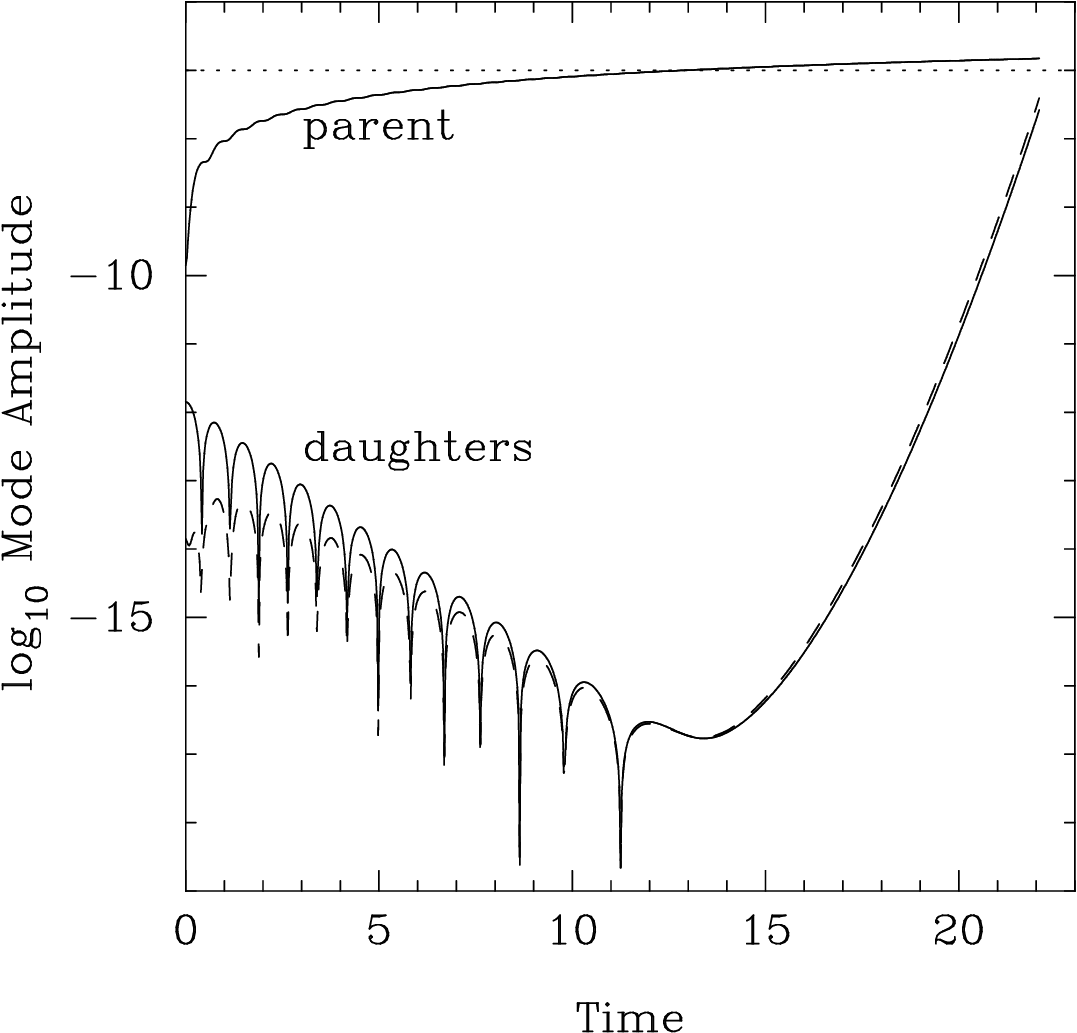}
\caption{Amplitude evolution of a three-wave system in which a parent mode $a$ (upper solid line) is linearly driven at a frequency $\omega$ and is coupled to a pair of daughters $(b,c)$, where $b$ is a $p$-mode (dashed line) and $c$ is a $g$-mode (lower solid line). Time is shown in units of $2\pi/\omega$. The parameters of the system are $\kappa_{abc}=10^7$, $U_a=3\times10^{-9}$, and in units of $\omega$, $(\omega_a, \gamma_a) = (1.005, 0.01)$,  $(\omega_b, \gamma_b) = (10^4, 1.0)$,  $(\omega_c, \gamma_c) = (0.7, 0.3)$. The modes are given initial amplitudes of $\log |q|=-10, -14, -12$, respectively. As the parent approaches its linear amplitude of $2\times10^{-7}$, it crosses the amplitude $\kappa_{abc}^{-1}$ (dotted line), and the daughters, which initially decay, become unstable and grow at a rate $\Gamma_{bc}\simeq\omega_c |\kappa_{abc} q_a|$.}\label{fig:3wave}
\vspace{0.1cm}
\end{figure}
   
Here the amplitude equations are second order in time whereas in WAQB they are first order in time. This is because here we adopt a configuration space mode expansion of the form $\vec{\xi}=\sum_a q_a\vec{\xi}_a$, where $\vec{\xi}$ is the Lagrangian fluid displacement and the sum is over all modes.  WAQB instead adopt a phase space mode expansion of the form $(\vec{\xi},\dot{\vec{\xi}})=\sum_a q_a (\vec{\xi}_a,-i\omega_a \vec{\xi}_a)$, where the sum is over all modes \emph{and} their complex conjugate. Both forms of the amplitude equations are, of course, valid and describe the same physics (see Appendix C in \citealt{Schenk:02}). 

By contrast with the non-resonant case, resonant daughters ($\omega_b+\omega_c\simeq \omega$) are unstable if
\beq
\label{eq:stability_resonant}
|q_a| >  \left|\kappa_{abc}\right|^{-1}\sqrt{\frac{\gamma_b\gamma_c}{\omega_b\omega_c}}\left[1+\frac{\Delta^2}{\left(\gamma_b+\gamma_c\right)^2}\right]^{1/2},
\eeq
where $\Delta = \omega_b+\omega_c-\omega$ is the detuning (see, e.g., \citealt{Nishikawa:68, Wu:01}; WAQB). If $|\Delta|\la\gamma_{b,c}\ll \omega_{b,c}$, the factor multiplying $|\kappa_{abc}|^{-1}$ in the above equation is $\ll 1$. Nonetheless,  if the non-resonant $\kappa_{abc}$ is much larger than the resonant $\kappa_{abc}$, the non-resonant instability can have a lower amplitude threshold.  This is indeed the case in coalescing NS binaries: in \S~\ref{sec:ns_structure} we show that the equilibrium tide couples to a $(p,g)$ daughter pair with $\sim 10^5$ times greater strength than it couples to a resonant daughter pair (independent of orbital frequency). As a result, the equilibrium tide is unstable to the PGI but not the resonant parametric instability.

Note that the resonant instability criterion derived in WAQB (eq. [\ref{eq:stability_resonant}]) does not reduce to the non-resonant criterion (eq. [\ref{eq:stability_criterion}]) in the limit  $\Delta \rightarrow \omega_b$. In deriving the resonant criterion, WAQB only accounted for the sum over modes and not their complex conjugate; as a result, equation (\ref{eq:stability_resonant}) is only accurate for nearly resonant daughter modes $\omega_b+\omega_c\simeq \omega$.

We illustrate the PGI in Figure \ref{fig:3wave} by numerically solving equations (\ref{eq:amp_eqn1})--(\ref{eq:amp_eqn3}) for a non-resonant, unstable three-wave system. The modes are each given different initial conditions (amplitudes, phases, etc.). We find that the instability is not sensitive to the initial conditions and that the daughters quickly attain similar amplitudes $|q_b|\simeq |q_c|$ once the nonlinear coupling terms dominate the linear terms.

Near the instability threshold $|q_a|\sim |\kappa_{abc}|^{-1}$ and the daughters' linear terms are comparable in magnitude to their second-order nonlinear terms (left and right hand sides of eqs. [\ref{eq:amp_eqn2}] and [\ref{eq:amp_eqn3}]). One might therefore wonder whether higher-order terms, which we neglect, are in fact important near threshold (or even whether perturbation theory is breaking down). We show in  \S~\ref{sec:higher_order_pg} that this is not the case.

\begin{figure}[t!]
\epsscale{1.1}
\plotone{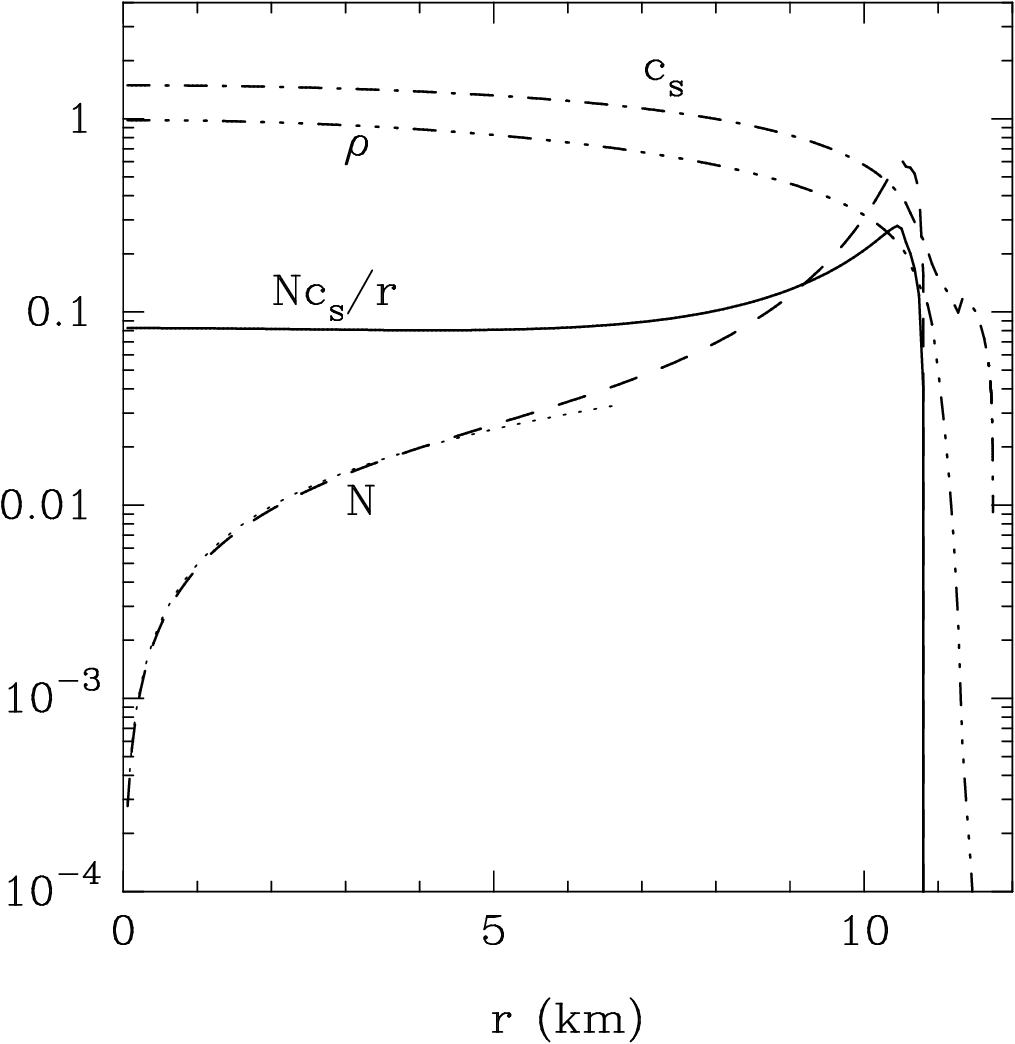}
\caption{Radial structure of an $M=1.4M_\odot$ NS assuming the SLy4 equation of state: the lines show $\rho$ (dash-triple-dot line; in units of $10^{15}\trm{ g cm}^{-3}$), $c_s$ (dash-dot line; in units of  $\omega_0 R$ where $\omega_0=(GM/R^3)^{1/2}$) , $N$ (dash line; in units of $\omega_0$), and $Nc_s/r$ (solid line; in units of $\omega_0^2$). The dotted line shows that $N/\omega_0\simeq 0.06 r/R$ for $r\la 5\trm{ km}$.}\label{fig:ns_structure}
\end{figure}
 
\section{Strength of  \lowercase{$p$}-\lowercase{$g$} mode coupling}
\label{sec:strength}

\begin{figure}
\epsscale{1.1}
\plotone{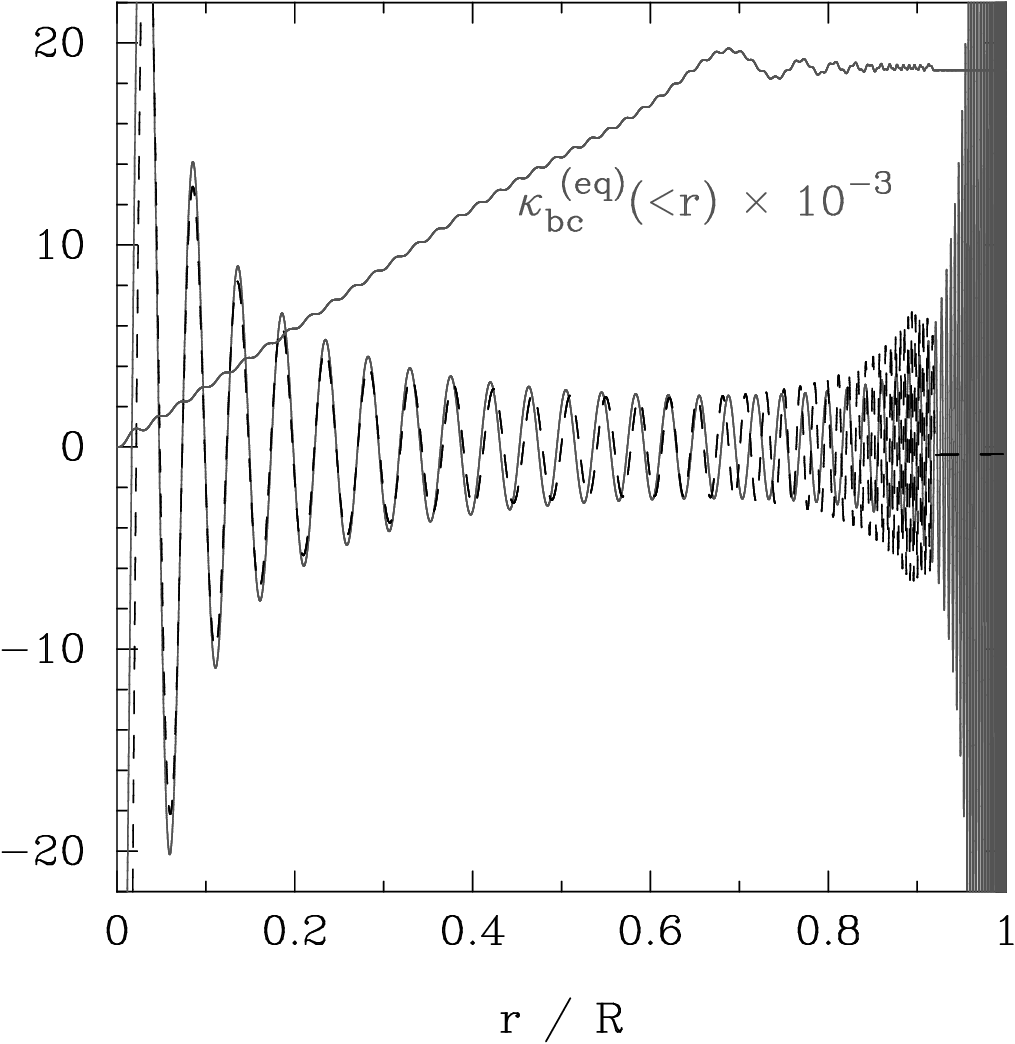}
\caption{Radial profile of the eigenfunctions of a $p$-mode (solid grey line) and a $g$-mode (dashed line) for the SLy4 NS model; their radial order, angular degree, azimuthal number, and frequency in units of $\omega_0$ are $(n_b,\ell_b, m_b, \omega_b)=(105, 1,-1, 200)$ and $(n_c, \ell_c, m_c, \omega_c)=(-80, 1,-1,6\times10^{-4})$.  The curves show $b_r \omega_b$ and $c_h \Lambda_c \omega_c$, respectively, in units of $R \omega_0$.  Note that the radial wavelengths of the modes are almost exactly equal for $r\la 0.6R$. The black line shows $\kappa_{bc}^{(\rm eq)}(<r)$ (multiplied by $10^{-3}$), the cumulative integral of the three-mode coupling coefficient between the $(p,g)$ pair and the $\ell=m=2$ equilibrium tide.}\label{fig:modes_kappa}
\vspace{0.2cm}
\end{figure}

The possible types of $p$-$g$ couplings include those with $(p,g)$ daughter pairs 
\[
g\co pg,  \hspace{0.4cm} p\co pg, \hspace{0.4cm} \trm{tide}\co pg,
\]
and those with $(p,p)$ or $(g,g)$ daughter pairs
\[
g\co pp, \hspace{0.4cm} p\co gg,\hspace{0.4cm}
\]
where $a\co bc$ denotes the coupling of a parent $a$ to a daughter pair $(b,c)$ with modes types $p$, $g$, or a tide (equilibrium or dynamical; we also show results for $f$-mode parents). First consider the couplings involving $(p,g)$ daughter pairs. As described in \S~\ref{sec:intro}, the constraint $|k_b-k_c|\la k_a$ implies that their frequencies must satisfy $\omega_b \omega_c\simeq \Lambda_c N c_s / r$. The spatial extent of the coupling region, and therefore the magnitude of $\kappa_{abc}$, depends on how rapidly $Nc_s/r$ varies with radius.  In stellar cores, often $N\propto r$ and $c_s\approx\trm{ constant}$ and the modes couple well throughout the core.   The coupling can be particularly strong in stars with large cores, such as NSs and white dwarfs.  In Figure \ref{fig:ns_structure} we show the structure of an $M=1.4 M_\odot$ ($R\simeq11.7\trm{ km}$) NS assuming the Skyrme Lyon (SLy4) equation of state \citep{Chabanat:98, Steiner:09}.\footnote{The model, including the crust, was kindly provided to us by A. Steiner. The $p$-mode and core $g$-mode that we consider couple in a region deep below the crust (see Fig. \ref{fig:modes_kappa}). Our results are therefore not sensitive to the detailed properties of the crust and for simplicity we assume that the NS is a completely fluid body and that $N^2=0$ in the crust (the outer $\simeq 1\trm{ km}$ of the NS).} We find that $Nc_s/r$ is nearly constant over a large region ($0 < r \la0.6R$), which we argue in  \S~\ref{sec:eos} is a feature of any NS equation of state  (although the magnitude of $Nc_s/r$ will depend somewhat on the equation of state).

\begin{figure}
\epsscale{1.15}
\plotone{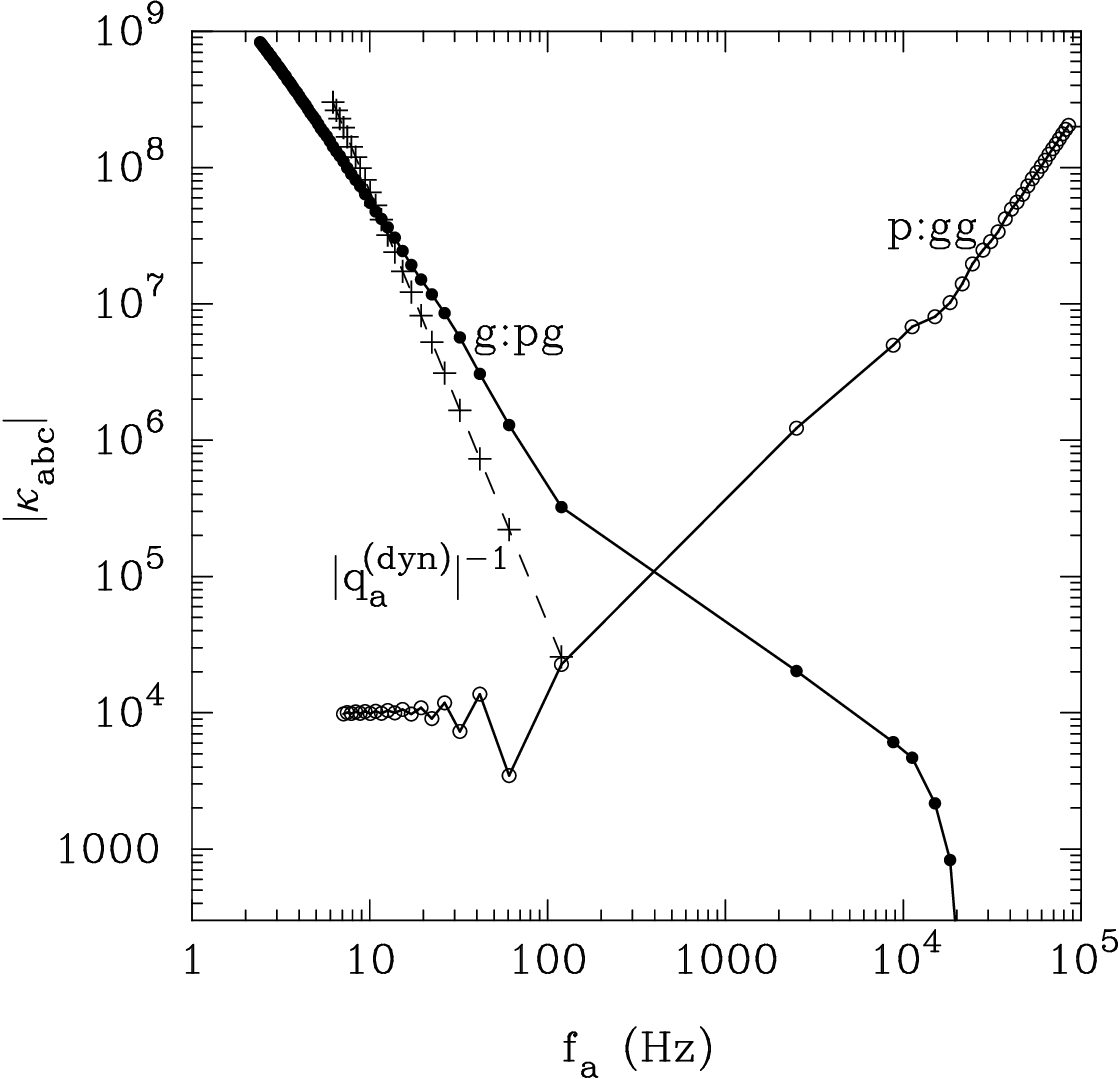}
\caption{Magnitude of the three mode coupling coefficient $\kappa_{abc}$ for a fixed daughter pair $(b,c)$ as a function of  parent mode eigenfrequency $f_a=\omega_a/2\pi$. The parents are $\ell_a=m_a=2$ modes and span the range from $g$-modes to $p$-modes (the $f$-mode is at $f_a=2.5\times10^3\trm{ Hz}$). The filled circles show $g\co pg$ coupling assuming the daughters used in Figure \ref{fig:modes_kappa}. The open circles show $p\co gg$ coupling for a self-coupled daughter that is the $g$-mode daughter used in Figure \ref{fig:modes_kappa}.  The points extend slightly into the regimes of $p\co pg$ and $g\co gg$ coupling. The crosses show the inverse of the dynamical tide amplitude $|q_{a}^{(\rm dyn)}|$ (see eq. [\ref{eq:qadyn}]). }\label{fig:kabc}
\end{figure}

A NS can support various types of oscillation modes, including $p$-modes and $g$-modes. The buoyancy that allows core $g$-modes to propagate is either thermally induced  \citep{McDermott:88, Gusakov:12} or due to proton-neutron composition gradients  \citep{Reisenegger:92}. Here we focus on the latter and assume a normal fluid NS. Note, however, that at temperatures $T\la 10^9\trm{ K}$, the bulk of the NS interior is expected to be a superfluid. The seismology of a superfluid NS can differ significantly from those of a normal fluid NS. In particular, the core of a \emph{zero temperature} superfluid NS does not appear to support propagating $g$-modes \citep{Lee:95, Andersson:01}. However, the core of a \emph{finite temperature} superfluid NS does support propagating $g$-modes \citep{Gusakov:12}, albeit with properties that are sensitive to the equation of state, the core temperature, and the model of nucleon superfluidity.  Given the uncertainties and difficulties associated with modeling $g$-modes in a finite temperature superfluid core, we assume, for simplicity, a normal fluid NS whose core $g$-modes are similar to those described in \citet{Reisenegger:92} and \citeauthor{Lai:94} (1994; and like these studies, we ignore general relativistic effects on the oscillations). Nonetheless, it is important to keep in mind that the results we present might be sensitive to superfluid effects. We briefly return to this topic in the conclusions (\S~\ref{sec:conclude}).

We also do not account for the rotation of the neutron star. Including rotation greatly complicates the analysis because instead of a single spherical harmonic, eigenfunctions must be expanded in a sum over spherical harmonics.   Rotation changes the properties of waves, especially those with mode frequency smaller than the rotation frequency. For the resonant parametric instability, rotation also lifts the frequency degeneracy relative to the azimuthal number $m$ (see, e.g., \citealt{Dziembowski:88}).  If the $p$-mode and $g$-mode have natural frequencies well above the rotation frequency, we do not expect their structure and therefore coupling strength to be significantly modified by rotation. 

 In Figure \ref{fig:modes_kappa} we show eigenfunctions of our NS model for a $(p,g)$ daughter pair that satisfy equation (\ref{eq:kb_eq_kc}) in the core. (We normalize the eigenvectors $\vec{a}$ as $\omega_a^2 \int d^3x \rho |\vec{a}|^2=E_0$, where $\rho$ is the stellar density and $E_0\equiv GM^2/R$.)  Because their wavelengths match over such a large region ($r\la0.6 R$), the coupling between the modes is very strong. In Figure \ref{fig:modes_kappa}, we show the coupling strength of this pair to the equilibrium tide, $\kappa_{bc}^{(\rm eq)}$, and in Figure \ref{fig:kabc} we show their coupling strength to an eigenmode parent, $\kappa_{abc}$, for parents that range from high-order $g$-modes to the $f$-mode to low-order $p$-modes (we calculate the coupling coefficients using the expressions given in WAQB; see also \S~\ref{sec:kappa_properties} below). Because the $p$-$g$ coupling occurs well-below the crust (which resides at $r> 0.9 R$), we do not expect it to be sensitive to the properties of the crust.

Since $\kappa_{abc}$ is symmetric in $abc$, if $g\co pg$ coupling can be strong then $p\co gg$ coupling can also be strong. For $p\co gg$ coupling, the constraint $|k_b-k_c|\la k_a$ implies that the daughter frequencies must satisfy $\Lambda_c \omega_b\simeq \Lambda_b \omega_c$ for small $k_a$ (unlike with a $(p,g)$ daughter pair, the daughter and parent wavelengths do not need to be equal).  Figure \ref{fig:kabc} shows $\kappa_{abc}$ for a self-coupled $g$-mode daughter coupled to a $p$-mode parent and demonstrates that $p\co gg$ coupling can indeed be strong.  While there may be stellar systems where $p\co gg$ coupling is important, for the remainder of the paper we focus on parent waves that are either $g$-modes or a tidal perturbation, and therefore we do not further consider $p\co gg$ coupling. We also do not consider the coupling of the equilibrium tide to the dynamical tide and a $p$-mode even though it is a form of $p$-$g$ coupling (since the dynamical tide is effectively a driven $g$-mode). WAQB showed that such coupling can lead to a linear-like steady-state driving of the $p$-mode (see WAQB's section 9 and Appendix B.3); it is therefore very different from the nonlinear exponential driving that characterizes the PGI.  Finally, for reasons given in \S~\ref{sec:kappa_properties}, we find that $g\co pp$ and $p\co pg$ coupling are much weaker than the other types of $p$-$g$ coupling.

We now describe the properties of the $p$-$g$ coupling coefficient in more detail. In \S~\ref{sec:kappa_properties} we present an analytic estimate of $\kappa_{abc}$, in \S~\ref{sec:ns_structure} we apply this estimate to the NS model, and in \S~\ref{sec:eos} we discuss the influence of the NS equation of state on the coupling strength.  In \S~\ref{sec:higher_order_pg} we explain why nonlinear terms beyond second-order in perturbation theory are not necessarily significant at the PGI threshold.

\subsection{Properties of $\kappa_{abc}$}
\label{sec:kappa_properties}
To see how $\kappa_{abc}$ depends on the stellar and mode parameters, assume that the daughters are short wavelength, low degree modes.  For a $(b,c)=(p,g)$ daughter pair, this implies that in the propagation regions $b_r \gg b_h$ and $c_r\ll c_h$, where the subscript $r$ and $h$ refer to the radial and horizontal displacement,
\bea
\label{eq:pmode_wkb}
(b_r, b_h)&\simeq& \frac{A_b}{\omega_b}\left(\cos\phi_b, \frac{c_s\sin\phi_b}{\omega_b r}\right),\\
\label{eq:gmode_wkb}
(c_r, c_h)&\simeq& \frac{A_c}{\omega_c}\left(\frac{\omega_c\sin\phi_c}{N}, \frac{\cos\phi_c}{\Lambda_c}\right),
\eea
$\phi_{b,c}\sim k_{b,c} r$ is the rapidly varying phase, $A_{b,c}=[E_0\alpha_{b,c}/\rho r^2]^{1/2}$, $\alpha_b=c_s^{-1}(\int c_s^{-1} dr)^{-1}$, and $\alpha_c=(N/r)(\int Nd\ln r)^{-1}$ (see, e.g., \citealt{Aerts:10}). An ordering of terms in WAQB's final expression for $\kappa_{abc}$ (lines A55-A62) reveals that their lines A60-A61 dominate and thus
\bea
\frac{d\kappa_{abc}}{d\ln r}&\simeq& \frac{\rho r^2}{E_0} F_b \omega_b^2 \left(a_r-a_h\right)b_r c_h\\
\label{eq:kappa_approx1}
&\simeq& \frac{F_b\left(\alpha_b\alpha_c\right)^{1/2}}{\Lambda_c} \left(a_r-a_h\right)\frac{\omega_b}{\omega_c}\cos\phi_b\cos\phi_c,
\eea
where $F_b$ is an angular integral that depends on the angular degree, $\ell$, and azimuthal number, $m$, of the three modes (see A21 in WAQB). Angular momentum conservation requires that the modes satisfy the selection rules $|\ell_b-\ell_c| \le \ell_a \le \ell_b+\ell_c$ with $\ell_a+\ell_b+\ell_c$ even and $m_a+m_b+m_c=0$. The angular factor $F_b/\Lambda_c$ is typically of order unity and depends only weakly on the $\ell$ and $m$ of the modes.

In order to obtain a form for $\kappa_{abc}$ that allows for accurate numerical integration, WAQB performed a series of integration by parts on the original, compact form for $\kappa_{abc}$ (see also \citealt{Wu:01}).  We took this numerically useful but non-compact form for $\kappa_{abc}$ as our starting point in deriving equation (\ref{eq:kappa_approx1}). However, for $p$-$g$ coupling, it is straightforward to derive this same equation starting from the original, compact form.  In Appendix  \ref{sec:app:alt_kap_derivation} we show this for the particular case of the equilibrium tide coupled to a $(p,g)$ daughter pair.

The coupling scales as $\omega_b/\omega_c\gg1$, the ratio of the $p$-mode to $g$-mode frequency.  If the radial wavelengths match, $\omega_b/\omega_c\simeq t_c/t_b$, where $t_{b,c}\simeq k_{b,c} r/\omega_{b,c}$ is the local radial group travel time of a mode.  Physically, the larger the ratio $\omega_b/\omega_c$ is, the longer the daughters (the $g$-mode, in particular) spend in the strong interaction region and therefore the larger $\kappa_{abc}$ is. 

The condition that the wavelengths match (eq. [\ref{eq:kb_eq_kc}]) implies that $\kappa_{abc}\propto \omega_b/\omega_c\propto \omega_b^2\propto\omega_c^{-2}$. In principle, $\kappa_{abc}$ can therefore be arbitrarily large.  However, if we require that the waves are global normal modes (i.e., standing waves), there is an upper limit, $\omega_{b, \rm max}$, to the $p$-mode frequency and a lower limit, $\omega_{c, \rm min}$, to the $g$-mode frequency. Depending on details of the stellar structure, $\omega_{b,\rm max}$ might be set by, e.g., the acoustic cutoff frequency of the atmosphere or the critical wavenumber $k_b\propto\omega_b$ above which linear  damping near the stellar surface is so rapid that the mode does not reflect. Similarly, since $k_c\propto \omega_c^{-1}$, local damping might also determine $\omega_{c, \rm min}$. In \S~\ref{sec:ns_structure} we describe the physics that determines $\omega_{b,\rm max}$ and $\omega_{c,\rm min}$ for a NS.

For $g\co pp$ coupling,  $\kappa_{abc}\propto\omega_b^2 b_r^2$ (we took $b=c$ because the interaction is maximized for self-coupled daughters) while for $g\co pg$ coupling of equal wavelength daughters, $\kappa_{abc}\propto \omega_b^2 b_r c_h$ (see eq. [\ref{eq:kappa_approx1}]). Since $b_r/c_h\sim \omega_c/\omega_b$, $g\co pp$ coupling is weaker than $g\co pg$ coupling by the ratio of the $g$-mode to $p$-mode frequency; by symmetry, the same argument applies to $p\co pg$ coupling. Because they are so much weaker, we do not further consider $g\co pp$ and $p\co pg$ coupling.
 
\subsection{Neutron star core}
\label{sec:ns_structure}

As Figure \ref{fig:ns_structure} shows, for $r\la 0.5R$ of the SLy4 NS model, $\rho\approx 10^{15}\trm{ g cm}^{-3}$, $N/r\approx0.06\omega_0/R$, and $c_s \approx 1.5 R\omega_0$, where $\omega_0=(GM/R^3)^{1/2}\simeq1.07\times10^4\trm{ rad s}^{-1}$ is the dynamical frequency. Furthermore, we find $\alpha_b \approx 0.6/R$ and $\alpha_c\approx 0.4/R$.  By equation (\ref{eq:kb_eq_kc}), this implies that the $p$-mode and $g$-mode wavelengths are approximately equal if
\beq
\label{eq:kb_eq_kc_NS}
\omega_b \simeq \frac{\Lambda_c N c_s}{r\omega_c}\simeq 90\Lambda_c \left(\frac{10^{-3}\omega_0}{\omega_c}\right) \omega_0.
\eeq
This relation will inform our choice of $\omega_b/\omega_c$ in the estimates below.

For a high-order $g$-mode parent coupled to a $(p,g)$ daughter pair with matched wavelengths, the coupling occurs near the parent's inner turning point $r_a\simeq (\omega_a/0.06\omega_0)R$, i.e., the location where $\omega_a\simeq N$. This is because the product of all three waves is largest and nearly constant near $r_a$. From equations (\ref{eq:kappa_approx1}) and (\ref{eq:kb_eq_kc_NS}) we find that at $r_a$,
\bea
\label{eq:kap_abc_NS}
\left.\frac{d\kappa_{abc}}{d\ln r} \right|_{r_a}&\simeq&\frac{0.01F_b}{\Lambda_a \Lambda_c} \left(\frac{M}{\rho R^3}\right)^{1/2} \frac{\omega_b}{\omega_c}\left(\frac{\omega_0}{\omega_a}\right)^2\non
&\simeq &
5\times10^5 \left(\frac{\omega_b}{200\omega_0}\right)^2\left(\frac{f_a}{100\trm{ Hz}}\right)^{-2}.
\eea
This expression agrees well with the magnitude and $\omega_a^{-2}$ scaling of the full $\kappa_{abc}$ integration (see Figure \ref{fig:kabc}). 

If the parent is the equilibrium tide, 
\bea
a_r \simeq \frac{GM}{gR}\left(\frac{r}{R}\right)^{\ell_a},\hspace{0.4cm} a_h\simeq \left(\frac{\chi+2}{\Lambda_a^2}\right) a_r,
\eea
where $g$ is the gravity and $\chi = \ell_a-\partial\ln g/\partial \ln r$
(see equations A12 and A13 in WAQB; although we include the gravitational perturbation due to the equilibrium tide in the full numerical calculations,  we ignore it in the analytic estimates below because it is a small effect). In the NS core $g\simeq 4\pi G\rho r/3$ and for $\ell_a=2$ and equal wavelength daughters we find from equations (\ref{eq:kappa_approx1}) and (\ref{eq:kb_eq_kc_NS}) 
\beq
\label{eq:kap_eq_NS}
\frac{d\kappa_{bc}^{(\rm eq)}}{d\ln r}\simeq 3\times10^4 \left(\frac{\omega_b}{200\omega_0}\right)^2\frac{r}{R}.
\eeq
The coupling scales linearly with $r$ in the region where the daughter wavelengths match, in good agreement with the full $\kappa_{bc}^{(\rm eq)}$ calculation (see Figure \ref{fig:modes_kappa}).\footnote{WAQB showed that there is an additional, potentially important, contribution to $\kappa_{bc}^{(\rm eq)}$ from the linear inhomogeneous terms in the equations of motion (see their Appendix A). For a $(p,g)$ daughter pair, however, the homogeneous terms dominate the coupling.}  By contrast, we find that the coupling strength between the equilibrium tide and a self-coupled $g$-mode daughter $c$ that is resonant with the tide is only $\kappa_{cc}^{(\rm eq)}\simeq 0.3$ (nearly independent of mode frequency and $\ell$ and $m$).\footnote{The arguments given in section 5.3 of WAQB explain why $\kappa_{cc}^{(\rm eq)}\simeq 0.3$. The coupling is weaker than that of $(p,g)$ daughter pairs by the factor $\trm{max}(\omega_b/\omega_c)\approx 10^5$ (see eq. [\ref{eq:max_omb_omc}]).  Note that we focus on self-coupled pairs because, as WAQB showed, the coupling strength peaks strongly for self-coupling.}

The coupling strength is near the maximum values given by  equations (\ref{eq:kap_abc_NS}) and (\ref{eq:kap_eq_NS}) as long as $|k_b -k_c|\la k_a$. Numerically, we find that high-order $p$-modes and $g$-modes satisfy the dispersion relations $\omega_b\simeq 2\omega_0 n_b$ and $\omega_c \simeq 0.035\omega_0 \Lambda_c / n_c$. It follows that within the NS core, $|k_b-k_c|\la 2\pi/R$ for $|n_b-n_c|$ less than a few; thus, since the lengthscale of the equilibrium tide is $\approx R$, for each $p$-mode, there are a few $g$-modes for which $\kappa_{bc}^{(\rm eq)}$ is close to the maximum value. As we describe in \S~\ref{sec:growth_rates},  one must account for this effect when determining how many modes are unstable to the PGI.

What is the maximum value of $\omega_b/\omega_c$ for which  equations (\ref{eq:kap_abc_NS}) and (\ref{eq:kap_eq_NS}) are valid? In \S~\ref{sec:local_driving} we show that the PGI requires that the $p$-modes (but not the $g$-modes) form standing waves.  We therefore argue that $\omega_{b, \rm max}$ is set by the acoustic cutoff frequency of the NS atmosphere $\omega_{\rm ac}\simeq c_s / 2H_\rho$,
where $H_{\rho}$ is the density scale height.  For $\omega_b>\omega_{\rm ac}$, acoustic waves do not reflect at the stellar surface and form standing waves.   In Appendix \ref{sec:app:acoustic_cutoff} we show that for a cold NS, $\omega_{\rm ac}\simeq 200 \rho_4^{-0.3} \omega_0$, where $\rho_4=\rho/10^{4}\trm{ g cm}^{-3}$.  At temperatures $T\sim 10^6\trm{ K}$ and densities $\rho \la 10^2\trm{ g cm}^{-3}$, ideal gas pressure dominates and $\omega_{\rm ac}\approx 1000\omega_0$.  We therefore expect the NS acoustic cutoff frequency to lie somewhere in the range $100\la \omega_{\rm ac}/\omega_0\la 1000$.  
  
Linear damping can also potentially limit the maximum value of $\omega_b/\omega_c$. In particular, if the $p$-mode linear damping rate exceeds the reciprocal of its round-trip travel time between turning points, $t_{\rm travel}\simeq2\int dr/c_s\simeq 2/\omega_0$, it will decay before reflecting and forming a standing wave (see, e.g., \citealt{Goodman:98}). In Appendix \ref{sec:app:acoustic_cutoff} we calculate the linear damping rate of modes and show that for $\omega_b<\omega_{\rm ac}$, the linear damping rate $\gamma_b \ll t_{\rm travel}^{-1}$. (Although the PGI does not require the $g$-mode to form a standing wave, we note that $\gamma_c < t_{\rm travel}^{-1}$ for low-degree $g$-modes with frequency $\omega_c\ga 10^{-3}\omega_0$, where here $ t_{\rm travel}\simeq2(\Lambda_c / \omega_c^2)\int N d\ln r$). Moreover, for low-degree modes that are unstable to the PGI (such as those shown in Figures  \ref{fig:modes_kappa} and \ref{fig:kabc}), we find $\gamma_{b,c} \ll \omega_{b,c}$ and thus the modes are well-described by the adiabatic stellar oscillation equations. We therefore conclude that for a $(p,g)$ daughter pair with nearly equal wavelengths (i.e., satisfying eq. [\ref{eq:kb_eq_kc_NS}]), the maximum value of $\omega_b/\omega_c$ is
\beq
\label{eq:max_omb_omc}
\trm{max}\left(\frac{\omega_b}{\omega_c}\right) \approx 4\times10^5\Lambda_c^{-1}\left(\frac{\omega_{\rm ac}}{200 \omega_0}\right)^2.
\eeq
This result motivates taking $\omega_b=200\omega_0$ as a reference value in equations (\ref{eq:kap_abc_NS}) and (\ref{eq:kap_eq_NS}).

\subsection{Influence of the equation of state on $\kappa_{abc}$}
\label{sec:eos}

We base our calculations of $\kappa_{abc}$ on a single type of nuclear equation of state (SLy4).  While SLy4 is consistent with all current observational constraints \citep{Steiner:10}, our choice is otherwise arbitrary; ideally we would like to compute $\kappa_{abc}$ for different equations of state.  Unfortunately, most microscopic calculations of high-density nuclear matter do not provide sufficient details to enable calculation of the buoyancy $N\propto c_s^2-c_e^2$ and therefore $\kappa_{abc}$.  In particular, they often provide the equilibrium sound speed $c_e$ but not the adiabatic sound speed $c_s$, where adiabatic in this context implies constant composition \citep{Reisenegger:92, Lai:94}.   

Although we do not have a precise estimate of $N$ for other equations of state, we believe that the results for SLy4 are representative.  \citet{Reisenegger:92} showed that in a NS core $N\approx (x/2)^{1/2} g/c_e$, where $x$ is the proton fraction.  Because $\rho$ is nearly constant over a large fraction of the core , $x^{1/2}$ and $c_e\simeq c_s$ change very little with $r$ while  $g\propto r$ (between the stellar center and $r\simeq R/2$). Therefore, for any equation of state, we expect $Nc_s/r$ to be nearly constant over a large portion of the star, i.e., similar to what is shown in Figure \ref{fig:ns_structure}. This implies that the $p$-mode and $g$-mode wavelengths can be equal over a large region, allowing for a strong nonlinear coupling.  Furthermore, based on the discussion in \citet{Reisenegger:92} and the SLy4 results, we expect the overall magnitude of $Nc_s/r$ to be within a factor of a few of the SLy4 value for any viable equation of state.\footnote{\citet{Lai:94} apply an approximate fitting method to extract $N$ from the four microscopic equations of state of \citet{Wiringa:88}. Figure 3 in \citeauthor{Lai:94} shows that all four yield high-order $g$-mode frequencies that are within a factor of two of each other for a given radial order. Moreover, three of the four have $g$-mode frequencies almost exactly equal to that of the SLy4 model (cf., the $g$-mode dispersion relation given in \S~\ref{sec:ns_structure}). Since $\omega_c\propto N$, this suggests that $N$ is likely within a factor of $\approx 2$ of the same value in all five equations of state.} Because $Nc_s/r$ determines the ratio $\omega_b/\omega_c$ and thus $\kappa_{abc}$ (see eqs. [\ref{eq:kappa_approx1}] and [\ref{eq:kb_eq_kc_NS}]), we conclude that most equations of state likely yield qualitatively similar $p$-$g$ coupling results.

\subsection{Higher-order $p$-$g$ coupling}
\label{sec:higher_order_pg}
As noted in \S~\ref{sec:stability}, at the PGI threshold the daughters' linear terms are comparable in magnitude to their lowest order ($n=2$) nonlinear terms.  Since the standard ordering of terms does not apply when going from $n=1$ to $n=2$, one might wonder whether it is valid to neglect $n>2$ terms. By carrying out a rough estimate of the magnitude of the $n=3$ terms, we now show that higher-order terms are unlikely to be important near threshold.

 At order $n=3$, the amplitude equation of a daughter mode $b$ includes four-wave coupling terms of the form $\kappa_{bdef} q_d^\ast q_e^\ast q_f^\ast$, where $\kappa_{bdef}$ is the four-wave coupling coefficient. The amplitude equations of the other modes include analogous four-wave terms. There are four types of four-wave couplings to consider: (i) $a \notin\{d,e,f\}$, (ii) $a=d\notin\{e,f\}$, (iii) $a=d=e\neq f$, (iv) $a=d=e=f$.  The last two cases correspond to a self-coupled parent. Because we are interested in assessing whether higher-order terms influence the onset of the instability, we assume the parent is at its initial, linear amplitude and the daughter amplitudes are infinitesimal (as in the stability analysis of \S~\ref{sec:stability}).

In cases (i) and (ii), the $n=3$ term in the amplitude equation contains the product of  three daughter amplitudes and two daughter amplitudes, respectively, whereas the $n=2$ term contains only a single daughter amplitude.  Since the daughter amplitudes are infinitesimal, the case (i) and (ii) $n=3$ terms  will necessarily be negligible compared to the $n=2$ terms at threshold; therefore they cannot prevent the onset of the $p$-$g$ instability (they might influence the saturation, however; see \S~\ref{sec:saturation}).

In case (iii), the $n=3$ term contains the product of two parent amplitudes $q_a^2$, whereas the $n=2$ term contains a single parent amplitude $q_a$.   However, the argument used in cases (i) and (ii) does not immediately apply because the parent amplitude is finite at threshold. Instead, to determine whether $n=3$ terms are important near threshold, we need to determine if $|q_a\kappa_{aabf}| > |\kappa_{abc}|$.  Note that the form of $\kappa_{abcd}$ is similar to that of $\kappa_{abc}$ (both are derived in \citealt{VanHoolst:94}); in particular, the integrand of $\kappa_{abc}$ contains terms of the form $(\grad\cdot \vec{\xi})^3$,  $\grad\cdot \vec{\xi} \xi^i_{;j}\xi^j_{;i}$, and $\xi^i_{;j}\xi^j_{;k}\xi^k_{;i}$ while the integrand of $\kappa_{abcd}$ contains terms of the form $(\grad\cdot \vec{\xi})^4$,  $(\grad\cdot \vec{\xi})^2 \xi^i_{;j}\xi^j_{;i}$, $(\xi^i_{;j}\xi^j_{;i})^2$, and $\xi^i_{;j}\xi^j_{;k}\xi^k_{;s}\xi^s_{;i}$. Here we use $\xi$ to represent the Lagrangian displacement of each of the modes (e.g., $(\grad\cdot \vec{\xi})^3\equiv \grad\cdot \vec{a}\grad\cdot \vec{b}\grad\cdot \vec{c}$), the subscript semi-colon denotes a covariant derivative, and we did not write down the terms involving the gravitational potential and its perturbations because they are negligible for the coupling of high-order modes.  Comparing forms, we see that the terms in $\kappa_{aabf}$ contain an extra factor of the parent's spatial derivative $\sim \partial a_i/\partial x^j$ relative to the terms in $\kappa_{abc}$.  Assuming $k_f\simeq k_b$ (otherwise $\kappa_{aabf}$ is negligible for the reasons given in \S~\ref{sec:intro}), then since $k_b\simeq k_c$ we see that the $n=3$ terms are important near threshold if $|q_a  \partial a_i/\partial x^j| \ga 1$ (note that this statement is independent of our choice of normalization). 

For a parent that is a low-order mode or the equilibrium tide, $\partial a_i/\partial x^j \la 1$.  Since we are interested in cases where $|q_a|\ll 1$ (e.g., for the equilibrium tide $|q_a|\sim (M'/M)(R/a)^3 \ll 1$), we have $|q_a  \partial a_i/\partial x^j| \ll 1$ and thus the $n=3$ term is negligible compared to the $n=2$ term near threshold. 

For a parent that is a high-order $g$-mode (e.g., a dynamical tide mode), $\partial a_i/\partial x^j \sim k_a a_r \gg 1$, where $k_a$ is the radial wavenumber of the parent. Thus, $n=3$ terms are comparable to $n=2$ terms if $|q_a k_a a_r|\ga 1$. This is just the usual nonlinearity parameter that determines the critical threshold above which a $g$-mode overturns the local stratification and breaks \citep{Goodman:98, Barker:10}.  If $\kappa_{abc} > \trm{max}(k_a a_r)$, there is a range of $|q_a|$ where $p$-$g$ coupling is unstable ($|\kappa_{abc} q_a| > 1$) but the $n=3$ term is negligible ($|q_a k_a a_r| < 1$). Using the values for the NS core and equation (\ref{eq:gmode_wkb}), we find $\trm{max}(k_a a_r) \approx 4\times 10^{-3} \Lambda_a (\omega_0/\omega_a)^3$, with the maximum occuring at the mode's inner turning point.    Combining this with our estimate of $\kappa_{abc}$ for $g\co pg$ coupling (eq. [\ref{eq:kap_abc_NS}]; see also Fig. \ref{fig:kabc}), we find that for $\omega_a / \omega_0 \ga 10^{-5}\Lambda_a$, there is a range of $|q_a|$ for which the parent is $p$-$g$ unstable but the $n=3$ terms are negligible. For the particular case of the dynamical tide in coalescing NS binaries, $|q_a k_a a_r|\ll 1$ at all frequencies $f_a$ (see eq. [\ref{eq:qadyn}] below) and therefore, as in the case of the equilibrium tide, the $n=3$ terms are negligible at the PGI threshold.

The analysis of case (iv) is similar to that of case (iii) except now the $n=3$ terms are important near threshold if $|q_a  \partial a_i/\partial x^j|^2 \ga 1$.  Since we found $|q_a  \partial a_i/\partial x^j|\ll 1$ in analyzing case (iii), the $n=3$ terms of case (iv) are also negligible near threshold.

Similar arguments apply to yet higher-order terms.  We therefore conclude that even though the magnitude of the $n=1$ and $n=2$ terms are similar at the $p$-$g$ instability threshold, the $n>2$ terms are not necessarily significant and a perturbative approach remains valid.

\section{Instability in coalescing neutron star binaries}
\label{sec:growth_rates}
 
In this section we consider the stability of the tide in coalescing NS-NS and NS-BH binaries. In \S\S~\ref{sec:NS_eqtide_growth_rate} and \ref{sec:NS_dyntide_growth_rate}  we determine when the equilibrium tide and dynamical tide are unstable to the PGI,  respectively. We then compute the nonlinear growth rates of the unstable daughters.  For comparison with the PGI, we also consider the stability of the tide to the resonant parametric instability. In \S~\ref{sec:well_above_threshold} we consider the implications if the parent is well-above the PGI threshold.  In \S~\ref{sec:local_driving} we show that the PGI growth rates are so large that the $g$-modes (but not the $p$-modes) grow significantly in less than their group travel time across the star, implying that the $g$-mode driving is local.

\subsection{Stability of the equilibrium tide}
\label{sec:NS_eqtide_growth_rate}
We derive the equilibrium tide $p$-$g$ stability criterion in Appendix \ref{sec:app:stability_analysis}. For a circular orbit, the dominant $\ell=2$ equilibrium tide is unstable to the PGI if  $\varepsilon \ga|\kappa_{bc}^{(\rm eq)}|^{-1}$ (cf. eq. [\ref{eq:stability_criterion}]), where the tidal amplitude factor 
\beq
\label{eq:epsilon_tide}
\varepsilon \simeq 
9\times10^{-4}\left(\frac{M'}{M_t}\right)\left(\frac{f_{\rm gw}}{100\trm{ Hz}}\right)^2.
\eeq
Here $M'$ is the companion mass, $M_t=M+M'$ is the total mass of the binary,  and $f_{\rm gw}$ is the gravitational wave frequency (which equals twice the orbital frequency).  From equation (\ref{eq:kap_eq_NS}), we thus find that the equilibrium tide is unstable to the PGI when
\beq
\label{eq:fi_eq}
f_{\rm gw} \ga 25\left(\frac{M'}{M_t}\right)^{-1/2} \left(\frac{\omega_b}{200\omega_0}\right)^{-1}\trm{ Hz}.
\eeq

When unstable, the equilibrium tide excites daughter pairs that grow exponentially at a rate
\bea
\Gamma_{bc}^{(\rm eq)}&\simeq& \omega_c| \varepsilon \kappa_{bc}^{(\rm eq)}|
\non &\simeq&
\label{eq:growth_rate_PGI_eq_tide}
75\left(\frac{M'}{M_t}\right) \left(\frac{\omega_b}{200\omega_0}\right)\left(\frac{f_{\rm gw}}{100\trm{ Hz}}\right)^2\trm{ Hz}.\hspace{0.4cm}
\eea
 This is much larger than the linear damping rate of $\ell=2$ modes in the frequency range of interest (see Appendix \ref{sec:app:acoustic_cutoff}). Moreover, it is much larger than the inverse of the gravitational wave inspiral timescale
\bea
t_{\rm gw}&\equiv& \frac{f_{\rm gw}}{\dot{f}_{\rm gw}}=
 5.9\left(\frac{\mathcal{M}}{1.2M_\odot}\right)^{-5/3}\left(\frac{f_{\rm gw}}{100\trm{ Hz}}\right)^{-8/3}\trm{ s}\hspace{0.5cm}
\eea
 \citep{Peters:63}, where the chirp mass $\mathcal{M}=[(MM')^{3}/M_t]^{1/5}$
 (for  $M=M'=1.4M_\odot$, $\mathcal{M} \simeq1.2M_\odot$).  The number of $e$-foldings that the daughters can grow before the binary merges is 
\bea
\int \Gamma_{bc}^{(\rm eq)} dt &=& \int \Gamma_{bc}^{(\rm eq)} \frac{df_{\rm gw}}{\dot{f}_{\rm gw}}
\non&\approx& 
10^3  \left(\frac{M^{3/5}M_t^{2/5}}{2.0M_\odot}\right)^{-5/3} \left(\frac{f_i}{25\trm{ Hz}}\right)^{-2/3},\hspace{0.5cm}
\eea
where $f_i$ is the value of $f_{\rm gw}$ when the daughters first become unstable as given by equation (\ref{eq:fi_eq}). Since the instability does not rely on resonant interactions, the daughters are continuously driven as the binary inspirals. The scaling $f^{-2/3}$ implies that most of the growth occurs at low frequencies, where the orbital decay is slowest. Because the number of $e$-foldings is so large, even daughters with very small initial amplitude can reach a significant amplitude well before the binary merges; the maximum amplitudes are therefore set by the nonlinear saturation of the instability rather than the time until merger.

The preceding calculation is for a single daughter pair.  However, since $n_b\simeq 100$ for $\omega_b\simeq 200\omega_0$ (see the $p$-mode dispersion relation given in \S~\ref{sec:ns_structure}), there are $\approx 10$ distinct $p$-modes that couple to the equilibrium tide with near equal effectiveness. Furthermore, as noted in \S~\ref{sec:ns_structure}, for each $p$-mode there are a few $g$-modes for which $\kappa_{bc}^{(\rm eq)}$ is near the maximum value. This suggests that for a given $(\ell_b, \ell_c)$, there are $\approx 10-100$ daughter pairs that have similarly large values of $\kappa_{bc}^{(\rm eq)}$.  And since the magnitude of $\kappa_{bc}^{(\rm eq)}$ is a weak function of the daughters' angular degree (it decreases only slightly with $\ell_{b,c}$), the total number can be greater still.  The number of daughters pairs that are $p$-$g$ unstable to the equilibrium tide can thus be $N_{\rm eq}\ga 100$. Moreover, the number increases as the orbit shrinks and  pairs with ever smaller $\kappa_{bc}^{(\rm eq)}$ become unstable. The net rate at which the PGI dissipates the equilibrium tide's energy might therefore be considerably larger than the rate due to a single daughter pair (see \S~\ref{sec:ns_Edot}).

We apply a similar analysis to evaluate the stability of the equilibrium tide to parametric resonance. This involves the coupling of the equilibrium tide to a pair of $g$-mode daughters with small detuning relative to the tidal frequency, $|\Delta|=|\omega_b+\omega_c - \omega| \ll \omega$ (for a circular orbit the tidal frequency $\omega$ equals twice the orbital frequency). Such daughters are unstable if their nonlinear growth rate $\Gamma_{bc} \ga \sqrt{\Delta^2 + \gamma_b\gamma_c}$, where $\Gamma_{bc}\approx \omega |\varepsilon \kappa_{bc}^{(\rm eq)}|$ (see WAQB).   We find that for self-coupled $g$-mode daughters  $\kappa_{bc}^{(\rm eq)}\simeq 0.3$, nearly independent of mode frequency and $\ell$ and $m$ (see \S~\ref{sec:ns_structure}). The parametric growth rate is therefore $\Gamma_{bc} \approx 10^{-3}(f_{\rm gw}/10\trm{ Hz})^3 \trm{ Hz}$ for an equal mass binary. While this is larger than $\gamma$ for resonant low $\ell$ modes with frequencies $\ga 10\trm{ Hz}$ (see Appendix \ref{sec:app:acoustic_cutoff}), it is smaller than their average detuning.\footnote{The $g$-mode dispersion relation given in \S~\ref{sec:ns_structure} determines the frequency spacing of the modes and implies that for self-coupling, $|\Delta|\simeq (10/\Lambda_a) (f_a/10\trm{ Hz})^2\trm{ rad s}^{-1}$. This is always greater than the resonant parametric growth rate $\Gamma_{bc}$ for low $\ell_a$ modes. As the binary inspirals and $\omega$ increases, $|\Delta|$ will be smaller than average for brief intervals.  However, because $\Gamma_{bc}$ is so small, the daughters will not have a chance to grow significantly during these brief intervals of instability.} We therefore conclude that the equilibrium tide is stable to  parametric resonance.

\subsection{Stability of the dynamical tide}
\label{sec:NS_dyntide_growth_rate}

The dynamical tide is unstable to the PGI if $|\kappa_{abc}| > |q_a^{(\rm dyn)}|^{-1}$ (cf. eq. [\ref{eq:stability_criterion}]), where $|q_a^{(\rm dyn)}|$ is the amplitude of the dynamical tide mode after it has undergone linear resonant driving. Following the calculation by \citeauthor{Lai:94} (1994; see also \citealt{Reisenegger:94b}), we find
\beq
\label{eq:qadyn}
|q_a^{(\rm dyn)}|\simeq 3\times10^{-5}h \left(\frac{f_a}{100 \trm{ Hz}}\right)^{19/6},
\eeq
where $h=(M'/M)^{1/2}(2M/M_t)^{5/6}$
and we made use of the expression for the linear overlap integral given in Appendix \ref{sec:app:acoustic_cutoff}  (note that our eigenfunction normalization is different than that of \citealt{Lai:94}). The mode frequency $f_a=\omega_a/2\pi$ equals the gravitational wave frequency $f_{\rm gw}$ when the resonance occurs. We verified this estimate of  $|q_a^{(\rm dyn)}|$ by numerically integrating the linear amplitude equations for an orbit decaying due to gravitational radiation.

Since the dynamical tide mode $a$ is a $g$-mode, it couples to a $(b,c)=(p,g)$ daughter pair with a strength $\kappa_{abc}$ given approximately by  equation (\ref{eq:kap_abc_NS}).  We thus find that the dynamical tide is unstable to the PGI if
\beq
\label{eq:fi_dyn}
f_a \ga 12 \, h^{-6/7}\left(\frac{\omega_b}{200\omega_0}\right)^{-12/7}\trm{ Hz},
\eeq  
in good agreement with the full numerical calculation shown in Figure \ref{fig:kabc}. The result depends only weakly on the mass ratio $M'/M$; e.g., for a NS-BH system with $M=1.4M_\odot$ and $M'=10M_\odot$, the dynamical tide is unstable for $f_a \ga 14\trm{ Hz}$.

When unstable, the dynamical tide mode excites daughter pairs that grow exponentially at a rate $\Gamma_{bc}^{(a, \rm dyn)}\simeq \omega_c |\kappa_{abc} q_a^{(\rm dyn)}|$, where the superscript $(a, \rm dyn)$ indicates that mode $a$ is a dynamical tide mode that was, at some earlier time, resonantly excited by the tide (see eq. [\ref{eq:growth_rate}]). From equations (\ref{eq:kap_abc_NS}) and (\ref{eq:qadyn}), we find
\beq
\label{eq:growth_rate_PGI_dyn_tide}
\Gamma_{bc}^{(a, \rm dyn)}\simeq 72 h \left(\frac{\omega_b}{200\omega_0}\right)\left(\frac{f_a}{100 \trm{ Hz}}\right)^{7/6}\trm{ Hz}.
\eeq
Like $\Gamma_{bc}^{(\rm eq)}$, the daughter growth rate $\Gamma_{bc}^{(a, \rm dyn)}$ is much larger than  $t_{\rm gw}^{-1}$ and the linear damping rate of relevant $\ell=2$ modes. The number of $e$-foldings that the daughters can grow between the time when they first become unstable at $f_{\rm gw}=f_i$ and the binary merges is
\beq
\int \Gamma_{bc}^{(\rm dyn)} dt \approx
10^3 h \left(\frac{\mathcal{M}}{1.2M_\odot}\right)^{-5/3}\left(\frac{f_i}{20\trm{ Hz}}\right)^{-3/2},
\eeq
where we used a value for $f_i$ motivated by equation (\ref{eq:fi_dyn}). This estimate assumes that nonlinear interactions do not prevent the parent from reaching the amplitude  $|q_a^{(\rm dyn)}|$ in the first place. Whether this is true depends on the initial amplitude of the daughters and the duration of the parent's linear excitation; if it is not true, the growth rate and the number of $e$-foldings will be smaller. 

As in the case of the equilibrium tide, there is enough time for many $e$-foldings of growth before the binary merges, with most of the growth occurring at low frequencies.  And like the equilibrium tide, there can be many daughter pairs that are unstable to the dynamical tide, i.e., $N_{\rm dyn} \gg 1$. A key difference, however, is that the equilibrium tide is continuously driven as the binary inspirals whereas the dynamical tide modes are only driven significantly for a brief interval near their linear resonance. We discuss the consequences of this in \S~\ref{sec:saturation}.

We now evaluate the stability of the dynamical tide to parametric resonance.  The maximum growth rate of parametrically unstable daughters is $\Gamma_{bc}\approx \omega_a |\kappa_{abc} q_a^{(\rm dyn)}|$, where $\kappa_{abc}$ is the coupling coefficient for the dynamical tide mode $a$ coupled to pair of $g$-mode daughters $(b,c)$ with frequencies $\omega_b+\omega_c\simeq \omega_a$. The relevant parent frequency is $\omega_a$ rather than the tidal driving frequency $\omega$ because post-resonance, the dynamical tide mode oscillates at its natural frequency $\omega_a$ (\citealt{Lai:94}). For  $f_a \ga 10\trm{ Hz}$, we find $\kappa_{abc} \approx 100$ nearly independent of $f_a$. The growth rate is thus $\Gamma_{bc}\approx 5h(f_a/\trm{ 100 Hz})^{25/6}\trm{ Hz}$, which is larger than the linear damping rate of low $\ell$ daughters with frequency $\ga 10\trm{ Hz}$ (see Appendix \ref{sec:app:acoustic_cutoff})  and their minimum $|\Delta|$  (because the dynamical tide $\kappa_{abc}$ is large as long as $|n_b-n_c| \la n_a$, $|\Delta|$ can be much smaller than the self-coupled $|\Delta|$ used in \S~\ref{sec:NS_dyntide_growth_rate} to determine the stability of the equilibrium tide to parametric resonance). We therefore conclude that the dynamical tide is unstable to parametric resonance for $f_a\ga 10\trm{ Hz}$.\footnote{For $f_a\la 10\trm{ Hz}$ we find  $\kappa_{abc}\approx100(f_a/10\trm{ Hz})^{-2}$ (it increases as $f_a^{-2}$ because at low frequencies $a_r\propto r_a^{-2}\propto f_a^{-2}$, where $r_a$ is the inner turning point). Nonetheless, the dynamical tide is stable for $f_a\la 10\trm{ Hz}$  due to the strong frequency dependence of $|q_a^{(\rm dyn)}|$ and the linear damping rate $\gamma$.}  However, since the PGI has a much larger growth rate for $10\la f_a \la 200\trm{ Hz}$, it is more likely the dominant instability.

\subsection{Validity of the PGI analysis well above threshold}
\label{sec:well_above_threshold}

When the parent is well above the PGI threshold, the nonlinear $n=2$ forces that act on the daughters dominate their linear forces. One might therefore worry whether the linear relations (e.g., the eigenfrequency) that we use to analyze the $p$-$g$ coupling remain valid when the parent is well above threshold.  However, this is not in fact a concern.  Since the linear eigenfunctions form a complete basis, they can be used to construct any vector field within the star.  The mathematical way we solve the full nonlinear fluid equations is to expand in this complete basis, including all the terms that seem significant.  The solutions will then show if the mode responds with its linear eigenfrequency or at some other frequency.  Thus, even if the $n=2$ terms are much larger than the linear terms, as long as the neglected terms really are small (a concern we address in \S~\ref{sec:higher_order_pg}), our approach captures the physics correctly.\footnote{In \S~\ref{sec:saturation} we show that the orbital phase error and tidal heating due to the PGI occur primarily near the equilibrium tide instability threshold. Therefore, as it happens, the $n=2$ terms are only slightly larger than the linear terms during the most important stage of the instability.}

\subsection{Local driving}
\label{sec:local_driving}

The PGI growth rates (eqs. [\ref{eq:growth_rate_PGI_eq_tide}] and [\ref{eq:growth_rate_PGI_dyn_tide}]) assume that the daughters are global standing waves. If, however, a daughter's growth rate within some region is much larger than its inverse group travel time  across that region, the daughter undergoes runaway local growth. It is then more proper to treat the daughter as a traveling wave rather than a standing wave. Although this does not affect the stability criterion (near threshold the growth rates are necessarily small), as we describe in \S~\ref{sec:saturation}, it influences the saturation of the instability.

In Appendix \ref{sec:app:local_growth_rate} we show that the local growth rate of the PGI is 
\bea
\widehat{\Gamma}_{bc}(r)&\simeq&\omega_c \left[\alpha_b \alpha_c\right]^{-1/2}\left|q_a \frac{d\kappa_{abc}}{dr}\right|\\
\label{eq:local_growth_rate}
&\approx& \omega_b \frac{F_b}{\Lambda_c}\left|\frac{q_a(a_r-a_h)}{r}\right|,
\eea
where we use the symbol $\widehat{\Gamma}$ to indicate a local growth rate and the second line follows from equation (\ref{eq:kappa_approx1}).  Based on the results of  \S~\ref{sec:ns_structure}, we see that for the equilibrium tide the local growth rate is larger than the global growth rate by a factor of about $(R\sqrt{\alpha_b\alpha_c})^{-1}\approx 2$, nearly independent of radius (recall that for the equilibrium tide $q_a\rightarrow \varepsilon$). The local and global growth rates are similar because the coupling is approximately constant throughout the core, which extends out to $r\approx R/2$.  For the dynamical tide, the local rate is larger than the global rate by $\approx (r/R)^{-1}$ because the coupling is strongest at small radii where the dynamical tide peaks; the maximum occurs at the tide's inner turning point $r_a/R\approx 0.1 (f_a/10\trm{ Hz})$. 

For a $p$-mode, the radial group travel time in the core is $t_{b}=r/c_s\approx 60 (r/R)\mu s$ and we find
\bea
\widehat{\Gamma}_{bc}^{(\rm eq)} t_{b} &\approx& 0.03\left(\frac{\varepsilon}{10^{-3}}\right) \left(\frac{\omega_b}{200\omega_0}\right)\frac{r}{R}  \\
\widehat{\Gamma}_{bc}^{(a, \rm dyn)} t_{b} &\approx& 0.01h  \left(\frac{f_a}{100\trm{ Hz}}\right) ^{7/6}
\eea
where for the dynamical tide we evaluated the result at $r_a$.  Since $\widehat{\Gamma}_{bc} t_{b} \ll 1$ for both the equilibrium and dynamical tides, the $p$-mode driving is in the standing wave limit. By contrast, for a $g$-mode with wavelength equal to that of a $p$-mode, $t_{c}\approx t_b (\omega_b/\omega_c)\gg t_b$ (see \S~\ref{sec:kappa_properties}) and we find
\bea
\widehat{\Gamma}_{bc}^{(\rm eq)} t_{c} &\approx& 3\times10^3\left(\frac{\varepsilon}{10^{-3}}\right) \left(\frac{\omega_b}{200\omega_0}\right)^3\frac{r}{R}  \\
\widehat{\Gamma}_{bc}^{(a, \rm dyn)} t_{c} &\approx& 10^3h  \left(\frac{\omega_b}{200\omega_0}\right)^2\left(\frac{f_a}{100\trm{ Hz}}\right) ^{7/6}.
\eea
The $g$-mode driving is therefore well into the traveling wave limit for typical parameter values. This means that the $g$-mode grows significantly in the time it takes to propagate across a small fraction of the star (unlike the $p$-mode, which undergoes many reflections in that same time).

\section{Saturation of the $\lowercase{p}$-$\lowercase{g}$ instability}
\label{sec:saturation}

Having evaluated the stability of daughters to the PGI in the previous sections, we now consider their saturation. As we discuss in \S~\ref{sec:sat_est}, determining precisely how, and at what energy, the daughters saturate is a challenging calculation that is beyond the scope of this paper. Instead, we leave the saturation energy as a free parameter whose magnitude we attempt to constrain based on a few informed assumptions.

We then use this result to determine how the PGI might influence the orbital evolution and heating rate of coalescing NS binaries. Because tidal interactions transfer energy and angular momentum from the binary orbit to the NS, they modify the rate of inspiral. This induces an orbital phase error $\Delta \phi$ relative to that of two point masses that accumulates over the course of the inspiral.  Furthermore, the waves excited by the tide heat the NS interior and deposit angular momentum that can spin-up the star.  In linear tidal theory, these effects are determined by the amplitude and linear damping rate of the parents, i.e., of the equilibrium and dynamical tides (see, e.g., \citealt{Lai:94}). In nonlinear tidal theory, however,  the daughters provide the parents with an additional, amplitude dependent, source of dissipation. In \S~\ref{sec:ns_Edot}, we evaluate the nonlinear energy dissipation rate due to the PGI and from that result,  estimate $\Delta \phi$ in \S~\ref{sec:phase_error}, and the tidal heating and spin-up of the NS in \S~\ref{sec:tidal_heating}.

\subsection{Daughter saturation energy}
\label{sec:sat_est}

\begin{figure}
\epsscale{1.1}
\plotone{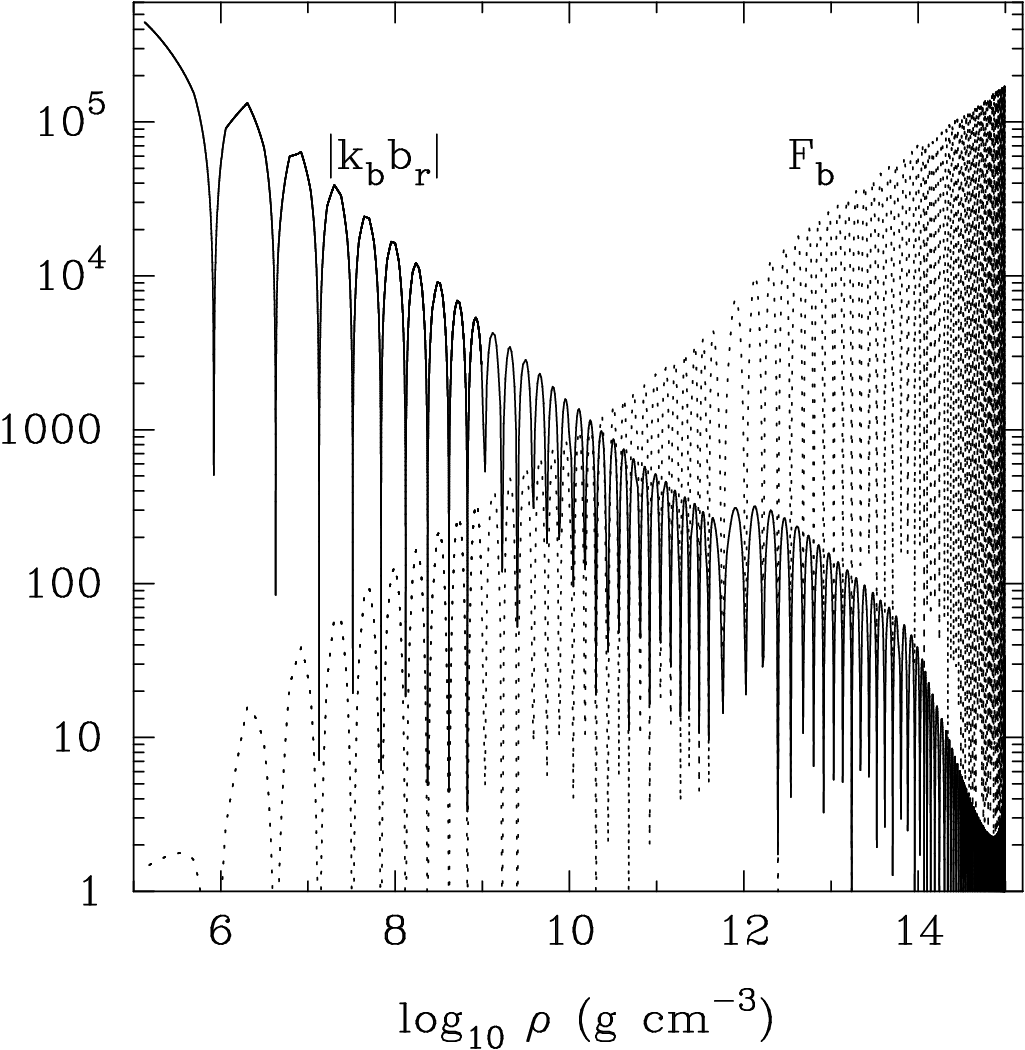}
\caption{Nonlinearity parameter $|k_b b_r|$ and energy flux $F_b$ for the $p$-mode used in Figure \ref{fig:modes_kappa} as a function of stellar density. The flux is in units of $10^7 E_0 \omega_0 R^{-2}$.}\label{fig:pmode_flux}
\end{figure}

The daughters' growth saturates when their nonlinear driving rate  balances their nonlinear damping rate. The latter depends on whether the saturation occurs in the ``discrete" limit or the ``continuum" limit (see \citealt{Arras:03}).  In the discrete limit, the daughters saturate by transferring their energy to a set of discrete modes whose linear damping rates are greater than the daughters' nonlinear driving rate. In the continuum limit, the daughters saturate by exciting a turbulent cascade in which the energy input at the outer scale (that of the parents) cascades down, via nonlinear interactions, to a very short wavelength inner scale where it rapidly thermalizes.

For coalescing NS binaries, the PGI growth rates are much larger than the linear damping rates of even the most highly damped, quasi-adiabatic $g$-modes (\S\S~\ref{sec:NS_eqtide_growth_rate} and \ref{sec:NS_dyntide_growth_rate}).  Moreover, since the driving of the daughter $g$-modes is local (\S~\ref{sec:local_driving}), their saturation must also occur locally. Together these suggest that for this problem, the saturation occurs in the continuum limit and involves the development of a turbulent cascade.   Because of the rapid local growth, the cascade might not saturate until the daughters become highly nonlinear and approach their local ``wave breaking" amplitude $|q_{\rm break}(r)|$.\footnote{It is not clear whether $|q_{\rm break}(r)|$ is a strict upper bound to the daughter amplitudes. Since the daughters oscillate at a frequency well-below their natural frequencies ($\sigma_b \ll \omega_b$ and $\sigma_c\ll \omega_c$; see Appendix \ref{sec:app:stability_analysis}), they might be able to reach amplitudes in excess of $|q_{\rm break}(r)|$.}  The daughter amplitudes probably cannot significantly exceed the linear amplitude of the parent $|q_a^{(\rm lin)}|$, however.  These considerations motivate us to parametrize the local saturation energy as
\beq
\label{eq:Esat_param}
E_{\rm sat}(r)\equiv\beta \min\left(|q_{\rm break}|^2, |q_a^{(\rm lin)}|^2\right) E_0,
\eeq
where the magnitude of the parameter $\beta$ is unknown but may be of order unity according to the above arguments.

Both $g$-modes and $p$-modes break when $|qk_r\xi_r| \sim 1$; at this amplitude, $g$-modes begin to overturn the stratification (see, e.g., \citealt{Goodman:98, Barker:10})  and $p$-modes impart order unity density perturbations. From equation (\ref{eq:gmode_wkb}), we find that $g$-modes whose wavelengths match their $p$-mode pair ($k_c\simeq k_b$) break at an amplitude 
\beq
\label{eq:qcbreak}
|q_{c, \rm break}(r)| \sim (k_c c_r)^{-1}\sim
10^{-4}\left(\frac{\omega_b}{200\omega_0}\right)^{-1}\left(\frac{r}{0.5R}\right)^{2}.
\eeq
For $p$-modes, $k_b b_r$ peaks near the stellar surface and for $\omega_b\approx 100\omega_0$ we find $(k_b b_r)^{-1}\sim 10^{-4}-10^{-5}$ near the outer turning point (see Figure \ref{fig:pmode_flux}). While this suggests that $|q_{b, \rm break}|\sim |q_{c, \rm break}|$, Figure \ref{fig:pmode_flux} shows that only a very small fraction of the $p$-mode energy flux $F_b=\omega_b^2 \rho r^2 b_r^2 c_s$ reaches the stellar surface (for a standing wave, its more proper to regard $F_b$ as a measure of the mode energy density rather than flux). As a result, wave breaking near the surface might not saturate the $p$-mode growth occurring deeper within the star. Instead, the $p$-mode might saturate only when it breaks in the core, which from equation (\ref{eq:pmode_wkb}) occurs at an amplitude (see also Figure \ref{fig:pmode_flux})
\beq
|q_{b,\rm break}(r)|\sim (k_b b_r)^{-1}
\sim 0.1 \left(\frac{r}{0.5R}\right).
\eeq
In that case, since $|q_{c, \rm break}|\ll |q_{b, \rm break}|$ at all radii, the $g$-mode always breaks before the $p$-mode (recall from \S~\ref{sec:stability} that $|q_b|\simeq |q_c|$ during the unstable growth). We will therefore assume that $|q_{\rm break}|=|q_{c, \rm break}|$.

 The amplitude of the equilibrium tide $\varepsilon$ is larger than $|q_{c,\rm break}|$ at all radii $r\la R/2$ when $f_{\rm gw}\ga 30(M'/M_t)^{-1/2}\trm{ Hz}$ (see eqs. [\ref{eq:epsilon_tide}] and [\ref{eq:qcbreak}]). There is therefore enough energy in the equilibrium tide at these frequencies that the $g$-mode daughters can, in principle, undergo local wave breaking throughout the core (i.e., reach $\beta\sim 1$ at all $r\la R/2$; recall that the equilibrium tide driving rate is nearly independent of radius). Since $|q_{c, \rm break}|\propto r^2$, the outer radii $r\simeq R/2$ would then be the principal seat of energy dissipation. Based on equation (\ref{eq:Esat_param}), we therefore express the volume integrated saturation energy of daughters driven by the equilibrium tide as
\bea
\label{eq:Esat_eq}
E_{\rm sat}^{(\rm eq)}&\equiv& \beta_{\rm eq} |q_{c, \rm break}(R/2)|^2E_0
\\ &\sim&
10^{-8} \beta_{\rm eq}\left(\frac{\omega_b}{200\omega_0}\right)^{-2} E_0,
\eea
where $\beta_{\rm eq}\sim 1$ might be a reasonable estimate.

For the dynamical tide, the driving occurs predominantly at the tide's inner turning radius $r_a/R\approx 0.1 (f_a/10\trm{ Hz})$. Plugging this radius into equation (\ref{eq:qcbreak}) and comparing the result with equation (\ref{eq:qadyn}), we find $|q_a^{(\rm dyn)}| \ll |q_{c, \rm break}|$  at all orbital frequencies; therefore, daughters driven by the dynamical tide do not break.  Instead, since the instability turns off once $|q_a|\la |\kappa_{abc}|^{-1}$, the daughters more likely saturate at an amplitude 
\beq
|q_a^{(\rm dyn)}|-|\kappa_{abc}|^{-1}\approx |q_a^{(\rm dyn)}|,
\eeq
i.e., after its resonant excitation, the dynamical tide mode $a$ transfers nearly all of its energy to the daughters. Based on equation (\ref{eq:Esat_param}), we therefore express the volume integrated saturation energy of daughters driven by the dynamical tide as
\bea
\label{eq:Esat_dyn}
E_{\rm sat}^{(a, \rm dyn)} &\equiv&\beta_{\rm dyn} |q_a^{(\rm dyn)}|^2E_0
\\ &\sim& 
10^{-9}\beta_{\rm dyn} h^2 \left(\frac{f_a}{100\trm{ Hz}}\right)^{19/3} E_0,
\eea
where $\beta_{\rm dyn}\sim 1$ might be a reasonable estimate.

\subsection{Nonlinear energy dissipation rate}
\label{sec:ns_Edot}

The nonlinear energy dissipation rate due to a single, saturated, daughter pair is $\dot{E}_{bc} \approx 2\Gamma_{bc} E_{\rm sat}$ (a factor of two appears because $\Gamma_{bc}$ is the growth rate of the amplitude not the energy).  
From equations (\ref{eq:growth_rate_PGI_eq_tide}) and (\ref{eq:Esat_eq}), the total dissipation rate due to the instability of the equilibrium tide is therefore
\bea
\label{eq:Edot_eq}
\dot{E}_{\rm eq}
&\sim&
 10^{48}  \beta_{\rm eq}N_{\rm eq}\left(\frac{M'}{M_t}\right)\left(\frac{\omega_b}{200\omega_0}\right)^{-1} 
\non &&\times
\left(\frac{f_{\rm gw}}{100\trm{ Hz}}\right)^2\trm{ erg s}^{-1},
\eea
where $N_{\rm eq}$ is the number of unstable daughters driven by the equilibrium tide (see \S~\ref{sec:NS_eqtide_growth_rate}) and we took a single value of $\omega_b$ as representative of all the unstable pairs. Similarly, from equations (\ref{eq:growth_rate_PGI_dyn_tide}) and (\ref{eq:Esat_dyn}), the total dissipation rate due to the instability of the dynamical tide is
\bea
\label{eq:Edot_dyn}
\dot{E}_{\rm dyn}
&\sim&
10^{47}\beta_{\rm dyn} N_{\rm dyn} h^3\left(\frac{\omega_b}{200\omega_0}\right) 
\non &&\times 
\left(\frac{f_a}{100\trm{ Hz}}\right)^{45/6}\trm{ erg s}^{-1}.\hspace{0.0cm}
\eea 
For a NS-NS binary at $f_{\rm gw}\simeq 100\trm{ Hz}$, these dissipation rates are $\sim 10^7N_{\rm eq}\beta_{\rm eq}$ and $\sim 10^4N_{\rm dyn}\beta_{\rm dyn}$ times larger than the linear dissipation rates of the equilibrium and dynamical tides, respectively (see \citealt{Lai:94}; this assumes a NS core temperature $T=10^8 K$).  Physically, this is possible because the PGI dissipates the energy in the tide on the rapid nonlinear driving timescale rather than the slow linear damping timescale.

Since each dynamical tide mode is excited for only a brief interval during its linear resonance, the mode ceases to heat the NS (i.e., $\dot{E}_{bc}^{(a, \rm dyn)}\rightarrow 0$) once the nonlinear interactions dissipate all of its energy. By contrast, since the fluid always seeks to follow gravitational equipotentials, there is a continuous flow of energy and angular momentum from the orbit into the equilibrium tide. Unstable daughters driven by the equilibrium tide therefore continuously dissipate energy at some finite rate $\dot{E}_{bc}^{(\rm eq)}$.

\subsection{Phase error $\Delta \phi$}
\label{sec:phase_error}

We now calculate the orbital phase error $\Delta \phi$ due to nonlinear tidal interactions. The total energy of the binary system is $E_{\rm tot}=E_{\rm orb}+E$, where $E_{\rm orb}$ is the orbital energy and $E$ is the energy in the modes and  the tidal interaction. Following a similar calculation by \citet{Lai:94}, the total power is
\beq
\dot{E}_{\rm tot}=\frac{dE_{\rm tot}}{df}\dot{f}=\left(\frac{dE_{\rm orb}}{df}+\frac{dE}{df}\right)\dot{f}
\eeq
and the change in phase over a time $dt$ is
\beq
\label{eq:dphi}
d\phi =2\pi f dt = 2\pi f \frac{df}{\dot{E}_{\rm tot}}\left(\frac{dE_{\rm orb}}{df}+\frac{dE}{df}\right).
\eeq
The first term in equation (\ref{eq:dphi}) is the point-mass result. The second terms represents the phase error due to the tidally induced change in stellar energy
\beq
\label{eq:dDphi}
d\left(\Delta \phi\right)=2\pi f \frac{df}{\dot{E}_{\rm tot}}
\frac{dE}{df}
\simeq 2\pi f \frac{df}{\dot{E}_{\rm gw}}
\frac{dE}{df}
= 2\pi t_{\rm gw} \frac{\dot{E}}{\dot{E}_{\rm gw}} df,
\eeq
where the gravitational wave luminosity
\beq
\dot{E}_{\rm gw} = 1.8\times10^{51} \left(\frac{\mathcal{M}}{1.2M_\odot}\right)^{10/3} \left(\frac{f_{\rm gw}}{100\trm{ Hz}}\right)^{10/3}\trm{ erg s}^{-1}
\eeq
and the second expression in equation (\ref{eq:dDphi}) follows because $\dot{E}_{\rm tot}\simeq \dot{E}_{\rm gw} \gg \dot{E}$, i.e., gravitational wave emission is the dominant source of orbital decay. 

For the equilibrium tide, $\dot{E}$ is given by equation (\ref{eq:Edot_eq}) and we find upon integrating equation (\ref{eq:dDphi})
\bea
\Delta \phi_{\rm eq} 
&\sim&
\beta_{\rm eq}N_{\rm eq}  \left(\frac{\mathcal{M}}{1.2M_\odot}\right)^{-5}\left(\frac{M'}{M_t}\right)
\non &&\times
\left(\frac{\omega_b}{200\omega_0}\right)^{-1} \left(\frac{f_i}{100\trm{ Hz}}\right)^{-3}\\
&\sim & 10\beta_{\rm eq}N_{\rm eq}  \left(\frac{\mathcal{M}}{1.2M_\odot}\right)^{-5}\left(\frac{M'}{M_t}\right)^{5/2}
\left(\frac{\omega_b}{200\omega_0}\right)^2, \non
\eea
 where in the second expression we used equation (\ref{eq:fi_eq}) to determine $f_i$, the value of $f_{\rm gw}$ at which daughters first become unstable. 
In \S~\ref{sec:NS_eqtide_growth_rate} we argued that $N_{\rm eq} \ga 100$ and possibly significantly larger.  Thus, even if $\beta_{\rm eq}\sim 10^{-3}-10^{-2}$ (i.e. considerably smaller than the upper bound set by energetic arguments), the instability of the equilibrium tide can induce a phase error $\Delta \phi_{\rm eq}> 1$. Moreover, since $\Delta \phi_{\rm eq}\propto f_i^{-3}$, the phase error accumulates primarily at large orbital separations.

As noted in \S~\ref{sec:ns_Edot}, nonlinear interactions do not affect the amount of orbital energy put into the dynamical tide mode, they only alter the rate at which the dynamical tide mode energy is thermalized (assuming again that the interactions do not influence the resonant excitation itself; see \S~\ref{sec:NS_dyntide_growth_rate}). As a result, even if the dynamical tide is unstable to the PGI, its phase error is the same as that of linear theory, i.e., $\Delta \phi_{\rm dyn} \ll 1$ \citep{Lai:94}.

\subsection{Tidal heating and spin-up of the neutron star}
\label{sec:tidal_heating}

The nonlinear dissipation rates $\dot{E}_{\rm eq}$ and $\dot{E}_{\rm dyn}$ (eqs. [\ref{eq:Edot_eq}] and [\ref{eq:Edot_dyn}]) describe how quickly orbital energy is converted to mode energy at saturation.  Assuming that the saturation is steady (at least in a time-averaged sense), the modes deposit their energy  into the star at a rate $\dot{E}=\dot{E}_{\rm eq}+\dot{E}_{\rm dyn}$. Some of this energy will heat the NS and some of it will go into rotational energy.  \citet{Bildsten:92} showed that the equilibrium tide cannot synchronize the spin of a standard NS. Although their calculation accounted only for the linear terms in the interaction Hamiltonian (for the nonlinear terms, see, e.g., WAQB), it is very unlikely that the PGI will synchronize the spin. This is because for spin frequencies $f_{\rm spin}\ga 20\trm{ Hz}$ the rotational energy $E_{\rm rot}\simeq 10^{51} (f_{\rm spin}/100 \trm{ Hz})^2\trm{ erg}$ is larger than the total energy removed from the orbit $\Delta E  \simeq \int \dot{E}_{\rm eq} dt\sim 10^{49} (f_i/100\trm{ Hz})^{-2/3}\trm{ erg}$, where we assumed uniform rotation and used equation (\ref{eq:Edot_eq}). From equation (\ref{eq:fi_eq}), $f_i\ga 20\trm{ Hz}$ and thus there is at most a small interval near the onset of the instability when the star can, in principle, briefly synchronize. 

While we conclude that synchronization is unlikely, without a detailed understanding of the saturation we do not know what fraction of $\dot{E}$ goes into heating the NS as opposed to spinning it up.  For simplicity, assume that all of $\dot{E}$ goes into heating the NS.  The thermal evolution of the NS heat content $U$ is then
\beq
\label{eq:dUdt}
\frac{dU}{dt}=\dot{E}_{\rm eq}+\dot{E}_{\rm dyn} + \dot{E}_{\rm cool}
\eeq
where
\beq
U\simeq 4.5\times10^{45} \left(\frac{T}{10^8 \trm{ K}}\right)^2\trm{ erg}
\eeq
\citep{Lai:94} and we neglect the cooling term $\dot{E}_{\rm cool}$ due to neutrino emission and surface photon emission which are small at temperatures $T\la 10^{10} \trm{ K}$ \citep{Meszaros:92}. 

Upon integrating equation (\ref{eq:dUdt}), we find that the equilibrium tide contribution alone yields a core temperature
\bea
T_{\rm eq}&\sim& 10^{10} \beta_{\rm eq}^{1/2}N_{\rm eq}^{1/2}  \left(\frac{\mathcal{M}}{1.2M_\odot}\right)^{-5/6}\left(\frac{M'}{M_t}\right)^{1/2}
\non &&\times
\left(\frac{\omega_b}{200\omega_0}\right)^{-1/2} \left(\frac{f_i}{100\trm{ Hz}}\right)^{-1/3} \trm{ K}.
\eea
For comparison, linear dissipation of the equilibrium tide in a NS-NS binary only heats the NS core to $T\approx 10^7 (f_{\rm gw}/100\trm{ Hz})^{5/6} \trm{ K}$ \citep{Lai:94}. 

The energy in the dynamical tide mode is rapidly thermalized due to nonlinear interactions. Its contribution to the core temperature is therefore found by solving $U(T_{\rm dyn})=|q_a^{(\rm dyn)}|^2E_0$, which yields 
$T_{\rm dyn}\sim 10^7 h  (f_a/100\trm{ Hz})^{19/6} \trm{ K}$. This is comparable to the linear result of \citet{Lai:94} because at frequencies $f_a\la 100\trm{ Hz}$, there is enough time for even linear damping to thermalize all the mode energy before the NS merges. We therefore conclude that $T_{\rm eq} \gg T_{\rm dyn}$.

\section{Summary and Conclusions}
\label{sec:conclude}

To summarize, we considered the nonresonant interaction between a parent mode and a pair of $(p, g)$ daughter modes. We first showed that if the parent mode amplitude $|q_a| > |\kappa_{abc}|^{-1}$, the system is unstable and the daughters grow at a rate $\Gamma_{bc}\simeq \omega_c|\kappa_{abc} q_a|$. We then evaluated the $p$-$g$ coupling coefficient $\kappa_{abc}$ and showed that it can be very large when the radial wavelength of the $p$-mode nearly equals that of the $g$-mode. 

After calculating $\kappa_{abc}$ for a NS model, we evaluated the stability of the tide in coalescing NS binaries to the PGI and the resonant parametric instability. We found that the equilibrium and dynamical tides are both unstable to the PGI at frequencies $f_{\rm gw}\ga 20\trm{ Hz}$ and excite daughter modes that grow $\sim 10^3$ times faster than the orbital inspiral rate $t_{\rm gw}^{-1}$. By contrast, resonant parametric coupling is either stable (in the case of the equilibrium tide) or excites daughters at a rate that is much smaller than the PGI growth rate (in the case of the dynamical tide).  

We then considered the saturation of the PGI in order to determine how it might influence a coalescing NS binary. Rather than attempt to solve for the saturation, which is a difficult problem, we left the saturation energy as a free parameter. We showed that if a daughter that is driven by the equilibrium tide saturates near its wave breaking amplitude, it induces an orbital phase error $\Delta \phi_{\rm eq} \ga 1$.  Since there are $N_{\rm eq}> 100$ unstable daughter modes, $\Delta \phi_{\rm eq}\gg 1$ is a possibility.  And because $\Delta \phi_{\rm eq}\propto f_{\rm gw}^{-3}$, most of the contribution to the phase error comes at large orbital separations ($f_{\rm gw} \la 50\trm{ Hz}$).  This may have important implications for ground based gravitational wave detections because the early inspiral contains an important portion of the signal  \citep{Cutler:93}. We also found that tidal heating due to the equilibrium tide PGI can raise the temperature of the NS core to $\sim 10^{10}\trm{ K}$ by $f_{\rm gw}\simeq 100\trm{ Hz}$. Such early heating of the NS might influence the electromagnetic and gravitational wave signature of the inspiral and merger. Finally, we found that the dynamical tide PGI does not significantly alter the linear theory estimates of $\Delta \phi_{\rm dyn}$ or dynamical tide heating found by \citet{Lai:94}. 

The saturation of the PGI is a significant source of uncertainty and if, instead, the daughters saturate at an amplitude well below their wave breaking amplitude, the instability will have only a small effect on the orbit and NS. Another important source of uncertainty that we briefly discussed is the influence of a superfluid core. Even though the core of a cold NS is expected to be superfluid, for simplicity we treated it as a normal fluid. Recently, \citet{Gusakov:12} showed that a superfluid core supports $g$-modes with a buoyancy frequency $N$ that can be much smaller than that of a normal fluid. Since the $p$-$g$ coupling coefficient $\kappa_{abc}\propto \max(\omega_b/\omega_c)\propto N^{-1}$, a superfluid core may have a larger $\kappa_{abc}$ and might therefore be unstable to the PGI at larger orbital separations (the local PGI growth rate is independent of $\omega_c$ and therefore $N$; see eq. [\ref{eq:local_growth_rate}]). Even so, the properties of $g$-modes in a superfluid core are  sensitive to the core temperature and equation of state and it is not clear to what extent $\kappa_{abc}$ changes in a superfluid. While a realistic treatment of the saturation and superfluidity present significant challenges, they are needed in order to accurately assess the importance of the PGI in coalescing NS binaries.

The PGI may be important in systems other than coalescing NS binaries. Two examples that we briefly mention are carbon/oxygen white dwarf binaries and binaries involving solar-type stars. Using a model of a $M=0.6M_\odot$ white dwarf, we find that a pair of $(p,g)$ daughter modes can couple to the equilibrium tide with a coupling strength $\kappa_{bc}^{(\rm eq)}\approx 10^3-10^4$.  This suggests that in equal mass carbon/oxygen white dwarf binaries, the equilibrium tide might be unstable to the PGI out to orbital periods $P\approx 15\trm{ min}$. In solar type stars, $(N/\omega_0)^2$ is much larger than in NSs and white dwarfs. As a result, a $p$-mode equals the wavelength of a $g$-mode (see eq. [\ref{eq:kb_eq_kc}])  only for $p$-mode frequencies well above the solar acoustic cutoff frequency ($\approx 60\omega_0$). Since the PGI requires $p$-modes that form standing waves, this suggests that solar-type stars are stable to this form of $p$-$g$ coupling.

Although we focused on the coupling of $(p,g)$ daughter pairs in a stellar core (where $Nc_s/r$ is nearly constant), there are other potentially interesting forms of $p$-$g$ coupling. For example, we showed that a high-order $p$-mode parent can couple strongly to $(g,g)$ daughter pairs. Such coupling, which does not require the $p$-mode wavelength to match the $g$-mode wavelength, might be important in systems where $p$-modes are driven to large amplitudes, e.g., by turbulent convection.   Another example is the coupling of $p$-modes to $g$-modes in the region near a radiative-convective boundary. Because a $g$-mode wavelength increases as it approaches a convection zone, at some point near the radiative-convective boundary the $g$-mode wavelength becomes long enough that it can equal the wavelength of a lower-order $p$-mode. While we find that such coupling is fairly weak near the base of the convective envelope of solar-type stars, it may be more significant in other types of stars, such as those that possess convective cores.

\acknowledgments We thank E. Quataert for useful discussions during the development of this work, D. Tsang for helpful comments on the original manuscript, and A. Steiner for providing equation of state and composition tables. This work was supported by NSF AST-0908873 and NASA  NNX09AF98G.

\begin{appendix}

\section{Stability analysis of  $\lowercase{p}$-$\lowercase{g}$ mode coupling}
\label{sec:app:stability_analysis}

In order to derive the $p$-$g$ instability criterion (eq. [\ref{eq:stability_criterion}]) and growth rate (eq. [\ref{eq:growth_rate}]), assume that the daughters are initially at infinitesimal amplitude and the parent is oscillating harmonically at a finite amplitude $q_a=|q_a| e^{-i\omega t}$. The daughter amplitude equations can then be written as
\bea
\label{eq:qb_ampeq}
\ddot{q}_b+\gamma_b \dot{q}_b +\omega_b^2q_b&=&\omega_b^2 K_{bc}^\ast q_c^\ast\\
\label{eq:qc_ampeq}
\ddot{q}_c+\gamma_c \dot{q}_c+\omega_c^2q_c&=&\omega_c^2 K_{bc}^\ast q_b^\ast,
\eea
where $K_{bc} \equiv K e^{-i\omega t}$ with $K= \kappa_{abc} q_a$. We will take mode $b$ to to be the $p$-mode and mode $c$ to be the $g$-mode.  First, to get a rough estimate of the instability criterion,  assume that  $\omega_b^2 q_b \gg \ddot{q}_b, \gamma_b \dot{q}_b$  (we will show that this is indeed the case when the daughters are unstable).  Then equation (\ref{eq:qb_ampeq}) implies $q_b\simeq K_{bc}^\ast q_c^\ast$ and by equation (\ref{eq:qc_ampeq}) we obtain
\beq
\ddot{q}_c+\gamma_c \dot{q}_c+\omega_c^2\left(1-|K|^2\right)q_c\simeq 0.
\eeq 
This is the equation of a damped oscillator and its solution is unstable if $|K| > 1$, independent of $\gamma_c$.  When unstable, the modes grow exponentially at a rate $\Gamma_{bc}\simeq \alpha\equiv\omega_c (|K|-1)$ (if $\alpha< \gamma_c/2$ then the growth rate is instead $\simeq |\alpha|^2/\gamma_c$;  however, for quasi-adiabatic modes, $\omega_c> \gamma_c$ and this slower growth applies only if $|K|$ is just barely larger than unity).  While growing, the $p$-mode oscillates harmonically at a rate $\omega$ whereas the $g$-mode mode does not oscillate at all.  Since $\omega$, $\gamma_b$, and $\omega_c |K|$ are all much less than $\omega_b$ for typical values, our initial assumption that $\omega_b^2 q_b \gg \ddot{q}_b, \gamma_b \dot{q}_b$ is approximately satisfied when the daughters are unstable. 

To evaluate the stability of the daughters more precisely (i.e., without assuming  $\omega_b^2 q_b \gg \ddot{q}_b, \gamma_b \dot{q}_b$) , let $q_{b,c}=A_{b,c}\exp(\theta_{b,c}t)$, where $A_{b,c}$ is a constant complex amplitude and $\theta_{b,c}=s+i\sigma_{b,c}$ with $s$ and $\sigma_{b,c}$ real constants. If we set $\sigma_b+\sigma_c=\omega$ (i.e., $\theta_b-\theta_c^\ast=i\omega$), the harmonic time dependence cancels and we find the characteristic equation
\beq
\label{eq:stab2}
\left[\theta_b^2+\gamma_b\theta_b+\omega_b^2\right]\left[(\theta_c^\ast)^2+\gamma_c \theta_c^\ast+\omega_c^2\right]=\omega_b^2\omega_c^2 |K|^2.
\eeq
The real (imaginary) part of this expression is a polynomial equation of order $s^4$ ($s^3$). Rather than explicitly solve these equations, we use the Routh-Hurwitz theorem to determine an approximate instability criterion. The theorem states that there exists an unstable root ($s>0$) if the order-zero term of a monic  polynomial is negative.  Applied to the real part of equation (\ref{eq:stab2}), we find that a \emph{sufficient} condition for instability is
\beq
\label{eq:stab3}
|K|^2 -1 > \frac{\sigma_b\sigma_c\left(\sigma_b\sigma_c + \gamma_b\gamma_c\right)}{\omega_b^2\omega_c^2}.
\eeq
For $p$-$g$ coupling of quasi-adiabatic modes $\omega_b \gg \omega_c, \omega$ and $\omega_{b,c} > \gamma_{b,c}$, and if $|K|\sim 1$ the real and imaginary parts of the characteristic equation imply $|\sigma_b \sigma_c|\simeq \omega^2(\omega_c/\omega_b)^2|K|^2\ll \omega^2$.  The instability condition (\ref{eq:stab3}) is therefore satisfied if $|K|\ga 1$ (i.e., the daughters are unstable if $|\kappa_{abc}| \ga |q_a|^{-1}$) and the characteristic equation shows that the instability growth rate is $\Gamma_{bc} \simeq \omega_c |\kappa_{abc} q_a|$. Moreover, $|\sigma_b|\simeq|\omega|$ and $|\sigma_c|\simeq |\omega |(\omega_c/\omega_b)^2$, and thus during the exponential growth 
the $p$-mode oscillates at nearly the forcing frequency $\omega$ while the $g$-mode oscillates at a frequency $\ll \omega, \omega_c$. We have confirmed these results with numerical experiments over a range of parameters.

When the parent is the  equilibrium tide, we can still express the stability criterion as $|K|>1$, where now (see WAQB)
\beq
K= \frac{M'}{M} \sum_{\ell m} W_{\ell m} X_k^{\ell m} \kappa_{bc}^{(\rm eq)} \left(\frac{R}{a}\right)^{\ell + 1}.
\eeq
Here $W_{\ell m} = 4\pi(2\ell + 1)^{-1}Y_{\ell m}(\pi/2, 0)$ and $X_k^{\ell m}$ is the Hansen coefficient for the $k$-th harmonic of the orbit, i.e., $\omega = k\Omega$.  For a circular orbit and $\ell=2$, $|K|\simeq \varepsilon |\kappa_{bc}^{(\rm eq)}|$, where the tidal amplitude factor $\varepsilon=(M'/M)(R/a)^3$.

\section{Alternative derivation of the equilibrium tide $\lowercase{p}$-$\lowercase{g}$ coupling coefficient} 
\label{sec:app:alt_kap_derivation}

In \S~\ref{sec:kappa_properties} we stated that for $p$-$g$ coupling it is straightforward to derive equation (\ref{eq:kappa_approx1}) starting from  the original, compact form for $\kappa_{abc}$.  Here we do so for the particular case of an equilibrium tide parent (mode $a$) coupled to a $p$-$g$ daughter pair $(b,c)$. In original, compact form (see, e.g., eq. [A24] in WAQB),
\bea
\label{eq:kappa_original}
 \kappa_{abc} &=& \frac{1}{E_0} \int d^3x \,
p 
\non && \times \bigg[ \left\{ \left(\Gamma_1-1\right)^2 + \frac{\partial \Gamma_1}{\partial
\ln \rho} \Big\rfloor_s \right\} \grad \cdot \vec{a} \grad \cdot
\vec{b} \grad \cdot \vec{c} 
\non && + a^i_{;j} b^j_{;k} c^k_{;i} + a^i_{;j} c^j_{;k} b^k_{;i}
\non && +\left(\Gamma_1-1\right) \left( a^i_{;j} b^j_{;i}
\grad \cdot \vec{c} + b^i_{;j} c^j_{;i} \grad \cdot \vec{a} + c^i_{;j}
a^j_{;i} \grad \cdot \vec{b} \right) 
\non && -\Gamma_1c_s^{-2} a^i b^j c^k \Phi_{;ijk}  \bigg], 
\eea
where the subscript semicolon denotes covariant derivative, $\Gamma_1$ is the adiabatic index, and $\Phi$ is the background gravitational potential. Because we are primarily interested in short wavelength perturbations or the equilibrium tide, this expression does not include the terms arising from perturbed gravity (i.e., unlike the full numerical calculation described in the main text, here we make the Cowling approximation).   In cartesian coordinates with $z$ pointing in the vertical direction, we have $\grad\cdot\vec{a}=0$ for the equilibrium tide,  $\grad\cdot\vec{b}\simeq \partial b_z/\partial z$ for the $p$-mode, and $\grad\cdot \vec{c}\simeq c_z/H \ll \partial c_z/\partial z \ll \partial c_x/\partial z$ for the $g$-mode (we assume, for simplicity, that the $p$- and $g$-modes do not vary in the $y$ direction). The dominant terms in equation (\ref{eq:kappa_original}) are therefore
\bea
 \kappa_{abc} &\simeq& \frac{1}{E_0} \int d^3x\,
p \left[a^i_{;j} c^j_{;k} b^k_{;i}
 +\left(\Gamma_1-1\right) c^i_{;j}a^j_{;i} \grad \cdot \vec{b} \right] 
 \non &\simeq& \frac{1}{E_0} \int d^3x\,
p\left[\frac{\partial a_z}{\partial x}\frac{\partial b_z}{\partial z} \frac{\partial c_x}{\partial z}+\left(\Gamma_1-1\right)\frac{\partial a_z}{\partial x}\frac{\partial b_z}{\partial z} \frac{\partial c_x}{\partial z}\right]
\non &\simeq& \frac{1}{E_0} \int d^3x\,
\Gamma_1 p \frac{\partial a_z}{\partial x}\frac{\partial b_z}{\partial z} \frac{\partial c_x}{\partial z}.
\eea
Since $\delta \rho_b/\rho \simeq -\partial b_z/\partial z$, the coupling depends on the compression due to the $p$-mode times the radial shear $\partial c_x/\partial z$ of the $g$-mode. The expressions for $b_z$ and $c_x$ are essentially the same as those given by $b_r$ and $c_h$ in equations (\ref{eq:pmode_wkb}) and (\ref{eq:gmode_wkb}), respectively, as long as the daughters' vertical wavelengths are much shorter than their horizontal wavelengths and the scale over which background quantities vary. Specifically, $b_z=A_{b} \cos \phi_b /\omega_b$, where $A_b=\left[E_0 \alpha_b/\rho S_{xy}\right]^{1/2}$, $\alpha_b=c_s^{-1}\left(\int c_s^{-1} dz\right)^{-1}$, and $S_{xy}\equiv \int dx dy$ is a constant horizontal area factor that cancels upon integration. Similarly,  $c_x=A_{c} \cos \phi_c /\omega_c$, where $A_c=\left[E_0 \alpha_c/\rho S_{xy}\right]^{1/2}$, and $\alpha_c=(N/z)\left(\int N d\ln z\right)^{-1}$. For daughters with nearly equal vertical wavelengths, $k_c\simeq k_b\simeq \omega_b/c_s$, and
\bea
\kappa_{abc}&\simeq& \frac{1}{E_0} \int d^3 x\,
\Gamma_1 p \frac{\partial a_z}{\partial x} \frac{ k_b^2 A_b A_c}{\omega_b \omega_c} \sin\phi_b\sin\phi_c 
\non &\simeq& 
 \frac{1}{E_0} \int d^3 x\,
\frac{\Gamma_1 p}{\rho c_s^2} \frac{E_0}{A_{xy}} \left[\alpha_b \alpha_c\right]^{1/2} \frac{\partial a_z}{\partial x} \frac{\omega_b}{\omega_c}  \sin\phi_b\sin\phi_c
\non &\simeq &
\int dz  \left[\alpha_b \alpha_c\right]^{1/2} \frac{\partial a_z}{\partial x} \frac{\omega_b}{\omega_c}  \sin\phi_b\sin\phi_c.
\eea
For the equilibrium tide $\partial a_z/\partial x \sim a_z /R$ and we see that this expression for the equilibrium tide $p$-$g$ coupling coefficient (i.e., $\kappa_{bc}^{(\rm eq)}$) agrees with equation (\ref{eq:kappa_approx1}) modulo the order unity constants of angular integration.

\section{Acoustic cutoff frequency, mode damping rate, and overlap integral}
\label{sec:app:acoustic_cutoff}

To estimate the NS acoustic cutoff frequency $\omega_{\rm ac}\simeq c_s/2H_\rho$, we use the \citet{Read:09}  parametrization of the low density EOS of cold matter (valid for the density range $\rho=[10^3, 10^{14}]\trm{ g cm}^{-3}$; see their Appendix C):
\bea
p(\rho) &=& K_i \rho^{\Gamma_i},\hspace{0.3cm}
\epsilon(\rho) = \rho c^2 + \frac{p}{\Gamma_i - 1},\hspace{0.3cm}
v_{\rm eq}^2(\rho) = \frac{\Gamma_i p c^2}{\epsilon + p},
\non 
\eea
where $\epsilon$ is the energy density, $v_{\rm eq} =\sqrt{dp/d\epsilon} \simeq c_s$ is the equilibrium sound speed, and the parameters $K_i$ and $\Gamma_i$ are step-functions of $\rho$ given in Table II of Read et al. For $\rho < 2\times10^7\trm{ g cm}^{-3}$, have $c_s \simeq 4.6\times10^7 \rho_4^{0.29}\trm{ cm s}^{-1}$ and 
\bea
\omega_{\rm ac}\simeq 210 \left(\frac{g_{14}}{2}\right) \rho_4^{-0.29} \omega_0,
\eea
where $\rho_4=\rho/10^4\trm{ g cm}^{-3}$ and $g_{14}=g/10^{14} \trm{ cm s}^{-2}$. For $\rho_2 \la \left[(T/\mu)/10^6\trm{ K}\right]^{1.71}$, where $\mu$ is the mean molecular weight, ideal gas pressure dominates and
\bea
\omega_{\rm ac}
&\simeq& 1400 \left(\frac{g_{14}}{2}\right)\left(\frac{T/\mu}{10^6\trm{ K}}\right)^{-1/2}\omega_0.
\eea
For simplicity, our calculation ignores the finite shear modulus of the NS crust and the density discontinuities therein. Since the energy density of the $p$-modes with frequency near $\omega_{\rm ac}$ is much larger in the core than in the crust (see Figure \ref{fig:pmode_flux}), we do not expect these properties of the crust to significantly affect the value of $\omega_{\rm ac}$.

The linear damping rate, $\gamma_a$, of a mode in a cold NS is dominated by viscous dissipation (see, e.g., \citealt{Reisenegger:92}). In order to calculate $\gamma_a$, we use the method described in \citet{Lai:94} and assume that the viscosity is dominated by electron-electron scattering as given by Lai's equation (8.23).\footnote{There appears to be an error in Lai's expressions for the shear and bulk damping rates (his equations 8.10 and 8.12): given his definition of $Y_{\ell m}$, there should not be a factor of $(\ell+|m|)!/(\ell-|m|)!$ multiplying the damping rates for modes with $m\neq 0$.  As a result, his expression 8.25 overestimates the damping rate of  $\ell=2, m=\pm2$ modes by a factor of 24.} For high-order modes, we find that for the $M=1.4M_\odot$ SLy4 NS model, the damping rates are well fit by the formula
\beq
\label{eq:linear_damping_rate}
\gamma_{a}\simeq\gamma_0 T_8^{-2} \Lambda_a^2  \left(\frac{\omega_a}{\omega_0}\right)^{\alpha},
\eeq
where $T_8=T/10^8\trm{ K}$ is the core temperature. For $p$-modes  $(\gamma_0, \alpha)=(2\times10^{-8} \trm{ s}^{-1}, 2)$ and for $g$-modes $(\gamma_0, \alpha)=(3\times10^{-9} \trm{ s}^{-1}, -2)$.

In \S~\ref{sec:NS_dyntide_growth_rate}, we  use the linear overlap integral
$I_{a\ell m}\equiv (MR^\ell)^{-1}\int d^3x\rho \vec{a}\cdot\grad(r^\ell Y_{\ell m})$ (see WAQB) to calculate the dynamical tide amplitude $|q_a^{(\rm dyn)}|$. For high-order $\ell=2$ modes, we find $I_{a\ell m}\simeq I_0 (\omega_a/\omega_0)^\beta$, where for $p$-modes $(I_0, \beta)=(10, -3)$ and for $g$-modes $(I_0, \beta)=(0.3, 2)$.

\section{Local growth rate of the $\lowercase{p}$-$\lowercase{g}$ instability}
\label{sec:app:local_growth_rate}
In order to derive the local growth rate of the PGI, we write the total Lagrangian displacement as a sum of the linear and nonlinear displacement $\vec{\xi}=\vec{\xi}_{\rm lin}+\vec{\xi}_{\rm nl}$. We can then obtain the equations of motion for $\vec{\xi}_{\rm nl}$ (see WAQB):
\bea
\rho \ddot{\vec{\xi}}_{\rm nl}=\vec{f}_1[\vec{\xi}_{\rm nl}]+2\vec{f}_2\left[\vec{\xi}_{\rm lin},\vec{\xi}_{\rm nl}\right]=\vec{f}_1[\vec{\xi}_{\rm nl}]+2\vec{f}_2\left[\vec{\xi}_{\rm lin}^\ast,\vec{\xi}_{\rm nl}^\ast\right],
\non
\eea
where $\vec{f}_1[\vec{\xi}]$ is the linear force and $\vec{f}_2[\vec{\xi},\vec{\xi}]$ is the leading-order nonlinear force, including the nonlinear tide term $\rho (\vec{\xi}_{\rm nl}\cdot\grad)\grad U$. The second equality, which follows because $\vec{\xi}$ is real, allows us to use WAQB's definition for $\kappa_{abc}$; namely, $\kappa_{abc}=2E_0^{-1}\int d^3x \vec{\xi}_a \cdot \vec{f}_2[\vec{\xi}_b,\vec{\xi}_c]$ (we absorb the factor of 2 multiplying $\vec{f}_2$ into our definition of the $p$-$g$ coupling coefficient).  Since we are interested in the growth rate of the PGI, we neglect the nonlinear inhomogeneous term $\vec{f}_2[\vec{\xi}_{\rm lin},\vec{\xi}_{\rm lin}]$ (see WAQB for a discussion of this term) and the saturation term  $\vec{f}_2[\vec{\xi}_{\rm nl},\vec{\xi}_{\rm nl}]$ (which is negligible when the daughter amplitudes are small). At small daughter amplitudes, the parent determines the linear displacement $\vec{\xi}_{\rm lin}=q_a\vec{\xi}_a$ and the $(b,c)=(p,g)$ daughter pair determine the nonlinear displacement $\vec{\xi}_{\rm nl}=q_b \vec{\xi}_b+q_c\vec{\xi}_c$. Therefore
\bea
\rho \left[\ddot{q}_b \vec{\xi}_b+\ddot{q}_c \vec{\xi}_c \right]&=&q_b\vec{f}_1[\vec{\xi}_b]+q_c\vec{f}_1[\vec{\xi}_c]+2q_a^\ast q_b^\ast\vec{f}_2\left[\vec{\xi}_a^\ast,\vec{\xi}_b^\ast\right]
\non && +2q_a^\ast q_c^\ast\vec{f}_2\left[\vec{\xi}_a^\ast,\vec{\xi}_c^\ast\right]
\non&=&
-\rho \omega_b^2 q_b\vec{\xi}_b-\rho\omega_c^2 q_c\vec{\xi}_c
+2q_a^\ast q_b^\ast\vec{f}_2\left[\vec{\xi}_a^\ast,\vec{\xi}_b^\ast\right]
\non && +2q_a^\ast q_c^\ast\vec{f}_2\left[\vec{\xi}_a^\ast,\vec{\xi}_c^\ast\right],
\eea
where we used the eigenvalue equation ${f}_1[\vec{\xi}_b]=-\rho \omega_b^2 \vec{\xi}_b$. Defining the radially-local inner product 
\beq
\langle \vec{\xi}_b,\vec{\xi}_c\rangle \equiv \int_r^{r+\Delta r} \int_{\Omega} d^3x \rho \vec{\xi}_b^\ast \cdot\vec{\xi}_c,
\eeq
where $\Delta r$ is many wavelengths long but $\ll r$ and $\Omega$ denotes the angular integral over the sphere, we find from equations (\ref{eq:pmode_wkb}) and (\ref{eq:gmode_wkb}) that for short wavelength daughters $\omega_b^2\langle \vec{\xi}_b,\vec{\xi}_b\rangle \simeq  E_0\alpha_b \Delta r$, $\omega_c^2\langle \vec{\xi}_c,\vec{\xi}_c\rangle \simeq  E_0\alpha_c \Delta r$, and $\langle \vec{\xi}_b,\vec{\xi}_c\rangle =0$ (the $p$-mode and $g$-mode are locally orthogonal even if their wavelengths match since, e.g., $b_r\propto \cos\phi_b$ and $c_r\propto \sin\phi_c$). Taking the local inner product of the equation of motion, we obtain the local amplitude equations for the daughters
\bea
\label{eq:local_amp1}
\ddot{q}_b+\omega_b^2q_b  &=&\omega_b^2 \alpha_b^{-1} \frac{d\kappa_{abc}^\ast}{dr}  q_a^\ast q_c^\ast\\
\label{eq:local_amp2}
\ddot{q}_c+\omega_c^2q_c  &=&\omega_c^2 \alpha_c^{-1} \frac{d\kappa_{abc}^\ast}{dr} q_a^\ast q_b^\ast,
\eea
where we neglected the self-coupling terms because they are small relative to the $p$-$g$ coupling terms and, since $\kappa_{abc}$ is a slowly varying function of $r$ in the coupling region, we took
\beq
\frac{1}{\Delta r} \int_r^{r+\Delta r} \int_{\Omega} d^3 x\frac{d\kappa_{abc}}{dr d\Omega}\simeq \frac{d\kappa_{abc}}{dr}.
\eeq
From the stability analysis of Appendix \ref{sec:app:stability_analysis}, we infer that when unstable, the local amplitude equations (\ref{eq:local_amp1}) and (\ref{eq:local_amp2}) yield a local daughter driving rate
\beq
\widehat{\Gamma}_{bc}\simeq \omega_c \left[\alpha_b \alpha_c\right]^{-1/2} \left| q_a\frac{d\kappa_{abc}}{dr}\right|.
\eeq
The global driving rate is related to the local driving rate by
\beq
\Gamma_{bc}\simeq \omega_c \left| q_a\kappa_{abc} \right| \simeq  \int dr \left[\alpha_b \alpha_c\right]^{1/2} \widehat{\Gamma}_{bc}.
\eeq
Since $\alpha_{b,c}=v_{b,c}^{-1}/\int v_{b,c}^{-1} dr$, where $v_b=c_s$ and $v_c=\omega_c^2 r /\Lambda_c N$ are the radial group velocities of the modes, this states that the global driving rate is the average of the local driving rate weighted by the geometric mean of the time the daughters spend at each radius.
\end{appendix}

\bibliographystyle{apj}

\vspace{0.5cm}
\bibliography{ref}

\begin{thebibliography}{45}
\expandafter\ifx\csname natexlab\endcsname\relax\def\natexlab#1{#1}\fi

\bibitem[{{Abadie} {et~al.}(2010){Abadie}, {Abbott}, {Abbott}, {Abernathy},
  {Accadia}, {Acernese}, {Adams}, {Adhikari}, {Ajith}, {Allen}, \&
  et~al.}]{LIGO:10}
{Abadie}, J., {et~al.} 2010, Classical and Quantum Gravity, 27, 173001

\bibitem[{{Aerts} {et~al.}(2010){Aerts}, {Christensen-Dalsgaard}, \&
  {Kurtz}}]{Aerts:10}
{Aerts}, C., {Christensen-Dalsgaard}, J., \& {Kurtz}, D.~W. 2010,
  {Asteroseismology (Berlin: Springer)}

\bibitem[{{Andersson} \& {Comer}(2001)}]{Andersson:01}
{Andersson}, N., \& {Comer}, G.~L. 2001, \mnras, 328, 1129

\bibitem[{{Arras} {et~al.}(2003){Arras}, {Flanagan}, {Morsink}, {Schenk},
  {Teukolsky}, \& {Wasserman}}]{Arras:03}
{Arras}, P., {Flanagan}, E.~E., {Morsink}, S.~M., {Schenk}, A.~K., {Teukolsky},
  S.~A., \& {Wasserman}, I. 2003, \apj, 591, 1129

\bibitem[{{Barker} \& {Ogilvie}(2010)}]{Barker:10}
{Barker}, A.~J., \& {Ogilvie}, G.~I. 2010, \mnras, 404, 1849

\bibitem[{{Bildsten} \& {Cutler}(1992)}]{Bildsten:92}
{Bildsten}, L., \& {Cutler}, C. 1992, \apj, 400, 175

\bibitem[{{Burkart} {et~al.}(2012){Burkart}, {Quataert}, {Arras}, \&
  {Weinberg}}]{Burkart:12}
{Burkart}, J., {Quataert}, E., {Arras}, P., \& {Weinberg}, N.~N. 2012, arXiv:
  1211.1393

\bibitem[{{Chabanat} {et~al.}(1998){Chabanat}, {Bonche}, {Haensel}, {Meyer}, \&
  {Schaeffer}}]{Chabanat:98}
{Chabanat}, E., {Bonche}, P., {Haensel}, P., {Meyer}, J., \& {Schaeffer}, R.
  1998, Nuclear Physics A, 635, 231

\bibitem[{{Cutler} {et~al.}(1993){Cutler}, {Apostolatos}, {Bildsten}, {Finn},
  {Flanagan}, {Kennefick}, {Markovic}, {Ori}, {Poisson}, \&
  {Sussman}}]{Cutler:93}
{Cutler}, C., {et~al.} 1993, Physical Review Letters, 70, 2984

\bibitem[{{Cutler} \& {Thorne}(2002)}]{Cutler:02}
{Cutler}, C., \& {Thorne}, K.~S. 2002, ArXiv General Relativity and Quantum
  Cosmology e-prints

\bibitem[{{Damour} {et~al.}(2012){Damour}, {Nagar}, \& {Villain}}]{Damour:12}
{Damour}, T., {Nagar}, A., \& {Villain}, L. 2012, \prd, 85, 123007

\bibitem[{{Dziembowski} {et~al.}(1988){Dziembowski}, {Krolikowska}, \&
  {Kosovichev}}]{Dziembowski:88}
{Dziembowski}, W., {Krolikowska}, M., \& {Kosovichev}, A. 1988, Acta
  Astronomica, 38, 61

\bibitem[{{Flanagan} \& {Hinderer}(2008)}]{Flanagan:08}
{Flanagan}, {\'E}.~{\'E}., \& {Hinderer}, T. 2008, \prd, 77, 021502

\bibitem[{{Flanagan} \& {Racine}(2007)}]{Flanagan:07}
{Flanagan}, {\'E}.~{\'E}., \& {Racine}, {\'E}. 2007, \prd, 75, 044001

\bibitem[{{Fuller} \& {Lai}(2012)}]{Fuller:12}
{Fuller}, J., \& {Lai}, D. 2012, \mnras, 421, 426

\bibitem[{{Goodman} \& {Dickson}(1998)}]{Goodman:98}
{Goodman}, J., \& {Dickson}, E.~S. 1998, \apj, 507, 938

\bibitem[{{Gusakov} \& {Kantor}(2012)}]{Gusakov:12}
{Gusakov}, M.~E., \& {Kantor}, E.~M. 2012, arXiv: 1211.4418

\bibitem[{{Hasselmann}(1967)}]{Hasselmann:67}
{Hasselmann}, K. 1967, Journal of Fluid Mechanics, 30, 737

\bibitem[{{Hinderer} {et~al.}(2010){Hinderer}, {Lackey}, {Lang}, \&
  {Read}}]{Hinderer:10}
{Hinderer}, T., {Lackey}, B.~D., {Lang}, R.~N., \& {Read}, J.~S. 2010, \prd,
  81, 123016

\bibitem[{{Ho} \& {Lai}(1999)}]{Ho:99}
{Ho}, W.~C.~G., \& {Lai}, D. 1999, \mnras, 308, 153

\bibitem[{{Kochanek}(1992)}]{Kochanek:92}
{Kochanek}, C.~S. 1992, \apj, 398, 234

\bibitem[{{Kumar} \& {Goldreich}(1989)}]{Kumar:89}
{Kumar}, P., \& {Goldreich}, P. 1989, \apj, 342, 558

\bibitem[{{Kumar} \& {Goodman}(1996)}]{Kumar:96}
{Kumar}, P., \& {Goodman}, J. 1996, \apj, 466, 946

\bibitem[{{Lai}(1994)}]{Lai:94}
{Lai}, D. 1994, \mnras, 270, 611

\bibitem[{{Lai} \& {Wu}(2006)}]{Lai:06}
{Lai}, D., \& {Wu}, Y. 2006, \prd, 74, 024007

\bibitem[{{Lee}(1995)}]{Lee:95}
{Lee}, U. 1995, \aap, 303, 515

\bibitem[{{McDermott} {et~al.}(1988){McDermott}, {van Horn}, \&
  {Hansen}}]{McDermott:88}
{McDermott}, P.~N., {van Horn}, H.~M., \& {Hansen}, C.~J. 1988, \apj, 325, 725

\bibitem[{{Meszaros} \& {Rees}(1992)}]{Meszaros:92}
{Meszaros}, P., \& {Rees}, M.~J. 1992, \apj, 397, 570

\bibitem[{{Moln{\'a}r} {et~al.}(2012){Moln{\'a}r}, {Koll{\'a}th}, {Szab{\'o}},
  {Bryson}, {Kolenberg}, {Mullally}, \& {Thompson}}]{Molnar:12}
{Moln{\'a}r}, L., {Koll{\'a}th}, Z., {Szab{\'o}}, R., {Bryson}, S.,
  {Kolenberg}, K., {Mullally}, F., \& {Thompson}, S.~E. 2012, \apjl, 757, L13

\bibitem[{{Nishikawa}(1968)}]{Nishikawa:68}
{Nishikawa}, K. 1968, Journal of the Physical Society of Japan, 24, 916

\bibitem[{{Oechslin} {et~al.}(2007){Oechslin}, {Janka}, \&
  {Marek}}]{Oechslin:07}
{Oechslin}, R., {Janka}, H.-T., \& {Marek}, A. 2007, \aap, 467, 395

\bibitem[{{Peters} \& {Mathews}(1963)}]{Peters:63}
{Peters}, P.~C., \& {Mathews}, J. 1963, Physical Review, 131, 435

\bibitem[{{Read} {et~al.}(2009){Read}, {Lackey}, {Owen}, \&
  {Friedman}}]{Read:09}
{Read}, J.~S., {Lackey}, B.~D., {Owen}, B.~J., \& {Friedman}, J.~L. 2009, \prd,
  79, 124032

\bibitem[{{Reisenegger} \& {Goldreich}(1992)}]{Reisenegger:92}
{Reisenegger}, A., \& {Goldreich}, P. 1992, \apj, 395, 240

\bibitem[{{Reisenegger} \& {Goldreich}(1994)}]{Reisenegger:94b}
---. 1994, \apj, 426, 688

\bibitem[{{Schenk} {et~al.}(2002){Schenk}, {Arras}, {Flanagan}, {Teukolsky}, \&
  {Wasserman}}]{Schenk:02}
{Schenk}, A.~K., {Arras}, P., {Flanagan}, {\'E}.~{\'E}., {Teukolsky}, S.~A., \&
  {Wasserman}, I. 2002, \prd, 65, 024001

\bibitem[{{Sekiguchi} {et~al.}(2011){Sekiguchi}, {Kiuchi}, {Kyutoku}, \&
  {Shibata}}]{Sekiguchi:11}
{Sekiguchi}, Y., {Kiuchi}, K., {Kyutoku}, K., \& {Shibata}, M. 2011, Physical
  Review Letters, 107, 051102

\bibitem[{{Steiner} {et~al.}(2010){Steiner}, {Lattimer}, \&
  {Brown}}]{Steiner:10}
{Steiner}, A.~W., {Lattimer}, J.~M., \& {Brown}, E.~F. 2010, \apj, 722, 33

\bibitem[{{Steiner} \& {Watts}(2009)}]{Steiner:09}
{Steiner}, A.~W., \& {Watts}, A.~L. 2009, Physical Review Letters, 103, 181101

\bibitem[{{Van Hoolst}(1994)}]{VanHoolst:94}
{Van Hoolst}, T. 1994, \aap, 286, 879

\bibitem[{{Weinberg} {et~al.}(2012){Weinberg}, {Arras}, {Quataert}, \&
  {Burkart}}]{Weinberg:12}
{Weinberg}, N.~N., {Arras}, P., {Quataert}, E., \& {Burkart}, J. 2012, \apj,
  751, 136 (WAQB)

\bibitem[{{Weinberg} \& {Quataert}(2008)}]{Weinberg:08}
{Weinberg}, N.~N., \& {Quataert}, E. 2008, \mnras, 387, L64

\bibitem[{{Wiringa} {et~al.}(1988){Wiringa}, {Fiks}, \&
  {Fabrocini}}]{Wiringa:88}
{Wiringa}, R.~B., {Fiks}, V., \& {Fabrocini}, A. 1988, \prc, 38, 1010

\bibitem[{{Wu}(1998)}]{Wu:98}
{Wu}, Y. 1998, PhD thesis, California Institute of Technology

\bibitem[{{Wu} \& {Goldreich}(2001)}]{Wu:01}
{Wu}, Y., \& {Goldreich}, P. 2001, \apj, 546, 469 (WG01)

\end{thebibliography}

\end{document}